\documentclass{article}
\usepackage{iclr2027_conference,times}
\usepackage{amsmath,amssymb,amsthm}
\usepackage{booktabs,multirow,graphicx,adjustbox,array,tabularx,float,longtable}
\usepackage[hidelinks]{hyperref}
\usepackage{url}
\usepackage{microtype}
\usepackage{xspace}
\usepackage{enumitem}
\usepackage{placeins}
\newtheorem{proposition}{Proposition}

\newcommand{\prstat}{PRR-Stat\xspace}
\newcommand{\prbudget}{PRR-B100\xspace}
\newcommand{\prr}{PRR\xspace}

\newcommand{\E}{\mathbb{E}}
\newcommand{\R}{\mathbb{R}}

\title{Which Histories Matter for Time Series Forecasting? Learning Predictive Relevance with Future Supervision}
\author{%
Yong-Hoon Choi \quad Kwang-Hyun Park \quad Youngjin Cho\\
Division of Robotics, Kwangwoon University\\
Seoul 01897, Republic of Korea\\
\texttt{yhchoi@kw.ac.kr \quad akaii@kw.ac.kr \quad yjaycho@kw.ac.kr}
}

\begin{document}
\raggedbottom
\iclrfinalcopy
\maketitle
\lhead{\footnotesize Which Histories Matter for Time Series Forecasting? Learning Predictive Relevance with Future Supervision}

\begin{abstract}
Retrieval-augmented time-series forecasting typically selects historical examples by similarity between observed pasts, although similar pasts can evolve differently. We propose \textbf{Predictive Relevance Retrieval (\prr)}, which uses realized future compatibility as privileged supervision to learn a retrieval function that remains strictly past-only at inference. \prr combines Pearson retrieval with a future-supervised predictive representation to expand candidate support, then reranks the union using statistical and learned pair relations. Across six datasets and four long horizons, \prr improves Pearson retrieval in \textbf{23 of 24 conditions}, reducing AnalogFutureMSE by \textbf{26.7\% on average}. A candidate-budget-matched variant, \prbudget, retains nearly the same retrieval improvement while using at most 100 candidates at inference, showing that the gain is not explained simply by a larger candidate pool. We then connect the retriever to five frozen forecasting backbones using validation-calibrated trust. Downstream effects are heterogeneous, and calibration primarily reduces harmful retrieval use rather than making improved retrieval universally beneficial. These results show that \textbf{predictive relevance and forecast utility are empirically distinct objectives}.
\end{abstract}

\section{Introduction}

Retrieval-augmented time-series forecasting uses historical observations as external predictive evidence: a system retrieves related past segments and exploits their observed futures when constructing a forecast \citep{tire2024raf,han2025raft,ning2025tsrag}. A natural retrieval baseline is Pearson correlation between observed past windows. Pearson is simple, training-free, and often effective, but it measures resemblance of the \emph{observed past}, not compatibility of the \emph{unseen future}. Histories that look alike can evolve differently under nonstationarity or regime change, motivating a more basic question: \emph{which histories should matter for the current query?}

We formulate this question as \emph{predictive relevance}. Realized futures, although unavailable at inference, retrospectively reveal which historical candidates were compatible with a training query. We use these futures as privileged supervision while requiring every deployed retrieval score to depend only on observed pasts. This yields \textbf{Predictive Relevance Retrieval (\prr)}, a framework that learns which candidates should be exposed and how they should be ranked. Pearson provides a conventional similarity view, a future-supervised temporal encoder provides a learned predictive view, and a fusion reranker ranks their combined candidate support using statistical and embedding relations.

Candidate support matters because reranking a fixed Pearson Top-$M$ list cannot recover a useful history that Pearson never exposes. \prr therefore retrieves Pearson Top-$M$ and learned-representation Top-$M$ histories independently from the same temporally admissible memory, unions the two sets, and applies future-supervised relevance ranking to the expanded pool. The query future is used only to construct supervision for historical training queries; candidate generation and reranking remain strictly past-only at validation and test time. For candidate-budget fairness, \prbudget reuses the same trained \prr model but truncates both retrieval branches before their union so that at most 100 candidates are reranked at inference. We retain \prstat as the more restricted fixed-support ablation, which reranks only the Pearson Top-$M$ support using the statistical pair representation.

Across six datasets and four long forecasting horizons, \prr improves Pearson retrieval in 23 of 24 conditions and reduces AnalogFutureMSE by 26.7\% on average. This establishes that future-supervised predictive relevance can improve upon Pearson correlation as a retrieval signal. The result, however, raises a distinct downstream question: \emph{does improved retrieval relevance relative to Pearson necessarily improve a strong direct forecaster?}

To study this question, we connect the same retriever to five frozen forecasting backbones---PatchTST \citep{nie2023patchtst}, iTransformer \citep{liu2024itransformer}, TimeMixer \citep{wang2024timemixer}, Seg-MoE \citep{ortigossa2026segmoe}, and DLinear \citep{zeng2023dlinear}. Retrieved and direct forecasts are combined using a bounded trust coefficient selected exclusively on validation data for each dataset--horizon--backbone condition; zero trust provides an explicit abstention to the Direct forecast. Across 120 matched conditions, validation-calibrated PRR improves or matches the frozen Direct forecast in 94 conditions, with a 1.41\% mean relative MSE reduction. Compared with common fixed trust, calibration substantially reduces harmful retrieval use, showing that predictive relevance and forecast trust are related but distinct problems.

The novelty is the retrieval formulation rather than its individual primitives: future outcomes supervise which histories are exposed and ordered while inference remains past-only.

Our contributions are fourfold. \textbf{(1)} We formulate historical retrieval as future-supervised predictive relevance, using realized futures as training-only privileged information while enforcing past-only inference. \textbf{(2)} We introduce \prr, which expands Pearson candidate support with a learned predictive representation and reranks the resulting union using observable statistical and representation relations. \textbf{(3)} On a matched 24-condition benchmark, \prr substantially improves Pearson; the \prbudget candidate-budget control retains nearly the same gain with at most 100 inference candidates, while the restricted \prstat fixed-support ablation and external retrieval diagnostics further test the source and strength of the improvement. \textbf{(4)} We connect retrieval relevance to five frozen forecasters over 120 matched conditions and show that validation-calibrated trust reduces harmful retrieval use, demonstrating that retrieval quality and forecast utility should be evaluated separately. Detailed fixed-support diagnostics, mechanism analyses, robustness studies, significance checks, and downstream trust analyses are provided in the Appendix.

\section{Related Work}

\paragraph{Retrieval-augmented time-series forecasting.}
Retrieval Augmented Forecasting (RAF) studies retrieval augmentation for time-series foundation models \citep{tire2024raf}, while RAFT retrieves historical segments with similar input patterns and exploits their observed futures \citep{han2025raft}. Subsequent work broadens both retrieval signals and aggregation mechanisms: TS-RAG uses learnable mixture-of-experts augmentation \citep{ning2025tsrag}; SARAF adapts relevance and diversity to stationarity \citep{zhou2026saraf}; Semantics-Enhanced Retrieval-Augmented Time Series Forecasting (SERAF) combines numerical and generated semantic retrieval \citep{zhou2026seraf}; Channel-wise Retrieval-Augmented Forecasting (CRAFT) performs channel-wise time- and frequency-domain retrieval \citep{kang2026craft}; and Cross-RAG selectively consumes retrieved references through cross-attention \citep{lee2026crossrag}. Predicting the Future by Retrieving the Past (PFRP) and Post-forecasting Identification and Revision (PIR) use global or local historical context to complement and revise direct forecasts \citep{du2026pfrp,liu2025pir}, while KReF interprets retrieved futures as a query-local empirical predictive distribution \citep{zhang2026kref}. Related nonparametric work such as kNN-MTS similarly gives forecasters direct access to historical nearest neighbors distributed across the dataset \citep{zhang2025knnmts}.

These methods establish the usefulness of historical memory, but typically design retrieval and aggregation in service of the final forecasting objective. Our question is more specific: \emph{what makes an individual historical example predictively relevant?} PRR uses realized future compatibility to supervise candidate retrieval itself, while requiring the deployed retriever to operate only on observed pasts. It also expands a conventional Pearson candidate set with a learned predictive-representation search before reranking their union. Thus PRR studies which histories should be exposed and ordered before downstream aggregation. Quantitatively, we compare isolatable retrieval rules (RAFT, SARAF, and CRAFT); methods coupling retrieval to semantic generation, adaptive mixing, or cross-attention consumption (e.g., TS-RAG, SERAF, and Cross-RAG) are treated as complementary forecasting systems rather than standalone retrievers.

\paragraph{Learning to rank and privileged supervision.}
Our reranker includes a listwise distributional objective, following the learning-to-rank view that the ranked list, rather than independent candidate pairs, can be the learning object \citep{cao2007listwise}. The distinction lies in the relevance signal and deployment constraint: the target ordering is induced by realized future compatibility, whereas the test-time score must be inferred entirely from past-observable information. Learned history retrieval has also been studied outside forecasting; for example, \citet{gammell2026history} learn input-dependent retrieval for nonstationary classification. More broadly, learning using privileged information allows variables available only during training to shape a predictor that must operate without them at test time \citep{vapnik2009lupi}. Privileged information has been used in time series to improve prediction from baseline variables \citep{karlsson2021privileged}, and TimeKD uses ground-truth future information in a teacher for privileged knowledge distillation \citep{liu2025timekd}. In PRR, by contrast, the privileged future neither becomes an inference input nor supervises the forecast directly; it defines which historical candidates should be considered relevant.

\paragraph{Nonstationarity and adaptation.}
Recent forecasting work also treats distribution shift and concept drift explicitly. DSOF combines fast and slow online adaptation streams without information leakage \citep{lau2025dsof}; proactive adaptation addresses concept drift under delayed future feedback \citep{zhao2025proactive}; LEAF separates longer-term macro-drift from shorter-term micro-drift \citep{chen2025leaf}; and TFPS uses pattern-specific experts for heterogeneous segment evolution \citep{sun2025tfps}. These approaches modify or adapt the forecasting model under change. Our setting is complementary: the direct forecaster remains frozen, allowing us to separate three questions---whether a history is predictively relevant, whether it adds information beyond the direct forecast, and whether its contribution should be trusted at inference.

\section{Predictive Relevance Retrieval with Future Supervision}
\subsection{Problem Formulation}
For a query past window $\mathbf{x}_q\in\R^L$, let $\mathbf{z}_q=\phi(\mathbf{x}_q)$ denote compact features derived only from the observed past, and define $\mathbf{o}_q=(\mathbf{x}_q,\mathbf{z}_q)$. Candidate $i$ is represented analogously by $\mathbf{o}_i$. All retrieval is restricted to a temporally admissible same-channel historical memory: the complete future of every candidate is already observed before the relevant query boundary. Our conventional baseline ranks candidates by Pearson correlation,
\begin{equation}
 s_{\mathrm{P}}(q,i)=\mathrm{corr}(\mathbf{x}_q,\mathbf{x}_i),
 \label{eq:pattern}
\end{equation}
implemented as cosine similarity after centering and $\ell_2$ normalization; for these univariate windows the two induce the same ranking.

Let $x_t^{\mathrm{train\text{-}norm}}$ denote a channel value normalized using training-period statistics. For a window ending at time $t$, its relative future trajectory is $y_t(h)=x_{t+h}^{\mathrm{train\text{-}norm}}-x_t^{\mathrm{train\text{-}norm}}$ for $h=1,\ldots,H$. Let $\mathbf{y}_q,\mathbf{y}_i\in\R^H$ denote these endpoint-anchored trajectories for query $q$ and candidate $i$. We define future distance and utility as
\begin{equation}
 d(q,i)=\frac{1}{H}\lVert\mathbf{y}_q-\mathbf{y}_i\rVert_2^2,
 \qquad u(q,i)=-d(q,i).
 \label{eq:utility}
\end{equation}
Since $\mathbf y_q$ is unavailable at inference, the population target for predictive retrieval is the utility recoverable from observable information,
\begin{equation}
 \boxed{r^*(\mathbf{o}_q,\mathbf{o}_i)=\E[u(q,i)\mid\mathbf{o}_q,\mathbf{o}_i].}
 \label{eq:bayes}
\end{equation}
Future trajectories therefore provide realized relevance labels only during training; no query future is available to the deployed retriever. This motivates the learning target but is not a finite-sample consistency guarantee.

\subsection{PRR: Learned Candidate Support and Fusion Reranking}
\prr uses two complementary past-only candidate views. The first is the conventional Pearson Top-$M$ set $\mathcal C^{\mathrm{P}}_M(q)$. The second is produced by a learned predictive representation. We use separate query and history encoders, $E_{\psi_q}$ and $E_{\psi_h}$, with the same architecture but distinct parameters. Each maps a normalized past window to a unit-norm 64-dimensional embedding. The implementation uses three 1D convolutional stages (64, 128, 128 channels; strides 2, 2, 2), GELU nonlinearities, adaptive average pooling, and a linear projection followed by $\ell_2$ normalization. Writing $\mathbf e_q=E_{\psi_q}(\mathbf x_q)$ and $\mathbf e_i=E_{\psi_h}(\mathbf x_i)$, learned similarity is
\begin{equation}
 s_{\mathrm{emb}}(q,i)=\mathbf e_q^\top\mathbf e_i.
 \label{eq:embsim}
\end{equation}

The embedding branch is trained from historical future compatibility. For a temporally valid training candidate list $\mathcal B(q)$, future distances are standardized within the list as $\widetilde d_{qi}=(d(q,i)-\mu_q)/\max(\sigma_q,10^{-6})$ and define a teacher distribution
\begin{equation}
 \pi_y(i\mid q)=\frac{\exp[-\widetilde d_{qi}/\tau_T]}{\sum_{j\in\mathcal B(q)}\exp[-\widetilde d_{qj}/\tau_T]},
\end{equation}
while past-only embedding similarities define a student distribution
\begin{equation}
 \pi_\psi(i\mid q)=\frac{\exp[s_{\mathrm{emb}}(q,i)/\tau_S]}{\sum_{j\in\mathcal B(q)}\exp[s_{\mathrm{emb}}(q,j)/\tau_S]}.
\end{equation}
The encoder pair minimizes $\mathrm{KL}(\pi_y\Vert\pi_\psi)$. Thus realized futures determine which historical relations the representation should preserve, but the learned similarity itself depends only on observed past windows. This representation-learning objective is distinct from the fusion-reranking objective introduced below; candidate sampling and temperatures are specified in Appendix~\ref{app:learned_support}.

For each query, \prr independently retrieves the Pearson and embedding Top-$M$ sets and forms
\begin{equation}
 \boxed{\mathcal C^{\mathrm{PRR}}(q)=\mathcal C^{\mathrm{P}}_M(q)\cup\mathcal C^{\mathrm{emb}}_M(q).}
 \label{eq:union}
\end{equation}
Hence $M\le |\mathcal C^{\mathrm{PRR}}(q)|\le 2M$, with overlap determined by the query. The fusion reranker then combines statistical and learned pair relations. The seven past-only statistics in $\phi$ are: current level relative to the full-window mean and standard deviation; final-quarter mean minus full-window mean, normalized by the full-window standard deviation; short-segment start-to-current change; full-window endpoint change; the ratio of short- to full-window first-difference volatility; normalized least-squares slope; and lag-1 autocorrelation. They form the statistical component of full \prr; using only this component defines the restricted \prstat ablation. The statistical representation is
\begin{equation}
 \mathbf f^{\mathrm{stat}}_{qi}=[s_{\mathrm{P}},\mathbf z_q,\mathbf z_i,\mathbf z_q-\mathbf z_i,|\mathbf z_q-\mathbf z_i|],
\end{equation}
and \prr augments it with learned query/candidate embeddings and their signed and absolute differences,
\begin{equation}
 \mathbf f^{\mathrm{PRR}}_{qi}=[\mathbf f^{\mathrm{stat}}_{qi},\mathbf e_q,\mathbf e_i,\mathbf e_q-\mathbf e_i,|\mathbf e_q-\mathbf e_i|]\in\R^{285}.
\end{equation}
A residual MLP with hidden width 384, LayerNorm, GELU, and dropout 0.1 predicts a correction $\Delta_\theta(\mathbf f^{\mathrm{PRR}}_{qi})$ to the Pearson score,
\begin{equation}
 \boxed{s_{\mathrm{PRR}}(q,i)=s_{\mathrm{P}}(q,i)+\alpha_r\Delta_\theta(\mathbf f^{\mathrm{PRR}}_{qi}),\qquad \alpha_r>0,}
 \label{eq:prrscore}
\end{equation}
where $\alpha_r$ is a learned positive residual scale. The Top-$K$ candidates under $s_{\mathrm{PRR}}$ form the retrieved analog set. Thus \prr changes both \emph{which candidates are visible} and \emph{which observable representation is available to rank them}.

\subsection{Future-Compatible Supervision and Past-Only Inference}
For a historical training query, future distances over the current PRR candidate list $\mathcal C^{\mathrm{PRR}}(q)$ are standardized by $\widetilde d_{qi}=(d(q,i)-\mu_q)/\max(\sigma_q,10^{-6})$ and converted to a soft relevance target
\begin{equation}
 p^*(i\mid q)=\frac{\exp(-\widetilde d_{qi}/\tau_y)}{\sum_{j\in\mathcal C^{\mathrm{PRR}}(q)}\exp(-\widetilde d_{qj}/\tau_y)}.
 \label{eq:target}
\end{equation}
For PRR scores in Eq.~\eqref{eq:prrscore}, the fusion reranker uses the listwise cross-entropy term
\begin{equation}
 \boxed{\mathcal L_{\mathrm{list}}=-\E_q\sum_{i\in\mathcal C^{\mathrm{PRR}}(q)} p^*(i\mid q)\log \mathrm{softmax}_i\,s_{\mathrm{PRR}}(q,i).}
 \label{eq:loss}
\end{equation}
The implemented objective adds a unit-weight Top-$K$ pairwise softplus term; Appendix~\ref{app:learned_support} gives the exact objective and training protocol. The query future never enters candidate generation or scoring at validation or test time. Candidate futures are already-observed historical memory: they define relevance targets during training and, at validation/test time, are accessed only after Top-$K$ selection to construct the historical retrieval forecast. This separation is the central deployment constraint of PRR: future information teaches relevance retrospectively but is absent from the retrieval decision for an unseen query.

\paragraph{Restricted fixed-support ablation.}
To isolate future-supervised reranking within fixed Pearson support, Appendix~\ref{app:fullresults} evaluates \prstat. It uses only the Pearson Top-$M$ set and the 29-dimensional statistical pair representation, with a two-layer residual MLP $s_{\mathrm{stat}}(q,i)=s_{\mathrm{P}}(q,i)+\alpha_r\Delta_\theta(\mathbf f^{\mathrm{stat}}_{qi})$. It is an ablation of \prr rather than a second main method.

\begin{figure}[t]
\centering
\includegraphics[width=\linewidth]{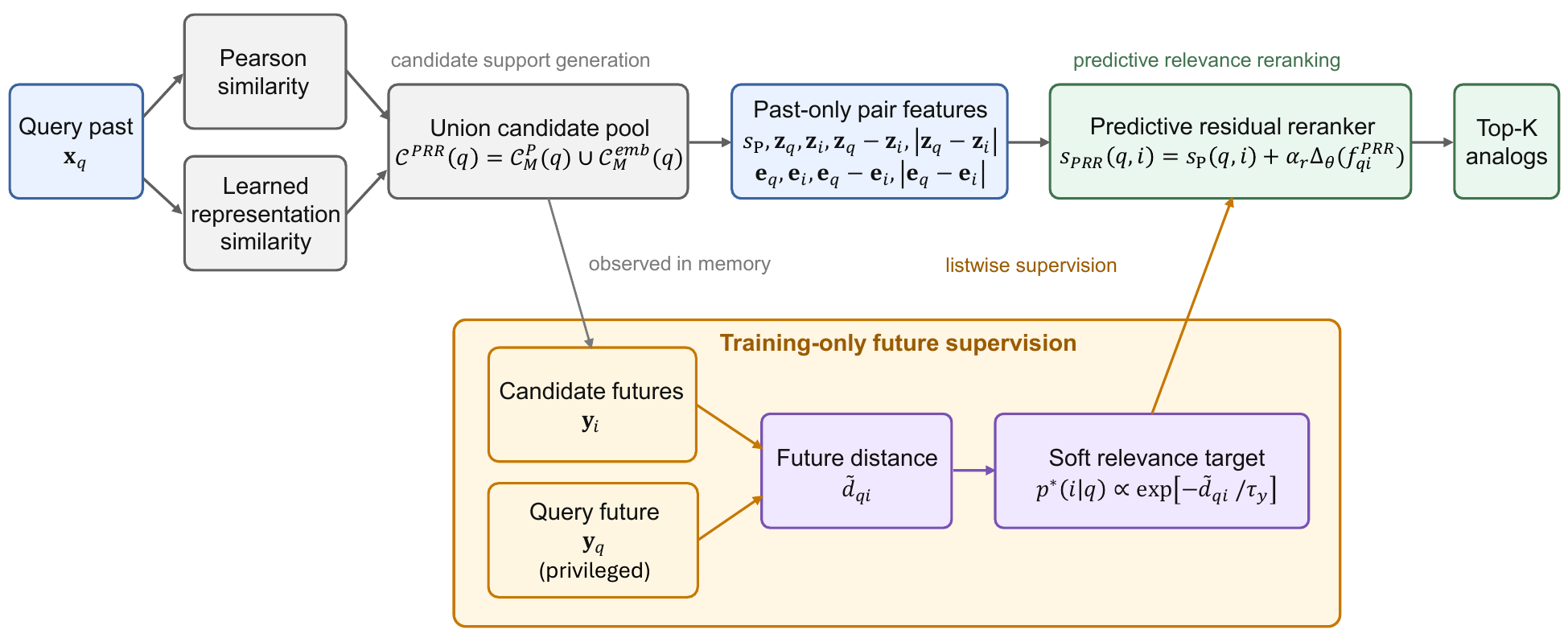}
\caption{\textbf{Overview of Predictive Relevance Retrieval (PRR).} Future compatibility provides training-only supervision; deployed retrieval remains strictly past-only.}
\label{fig:method}
\end{figure}

Figure~\ref{fig:method} summarizes this separation between training supervision and deployment: future compatibility supplies privileged targets during training, whereas candidate generation and reranking use only past-observable information at inference.

\section{Interpreting Predictive Relevance}
Equation~\eqref{eq:bayes} defines the observable relevance that \prr seeks to infer from past-only information. We now use this quantity to distinguish two conceptually different sources of historical usefulness: candidates that are broadly useful across many queries and candidates whose value depends on the current query. Formal population characterization and orthogonality results are given in Appendix~\ref{app:proofs}; the main text uses the decomposition only as an interpretive framework for the controlled diagnostics below.

\subsection{Candidate-Level Utility and Query-Specific Compatibility}
Define
\begin{equation}
G^*(\mathbf{o}_i)=\E[u(q,i)\mid\mathbf{o}_i],\qquad
H^*(\mathbf{o}_q,\mathbf{o}_i)=r^*(\mathbf{o}_q,\mathbf{o}_i)-G^*(\mathbf{o}_i).
\label{eq:ghdef}
\end{equation}
Then
\begin{equation}
\boxed{r^*(\mathbf{o}_q,\mathbf{o}_i)=G^*(\mathbf{o}_i)+H^*(\mathbf{o}_q,\mathbf{o}_i).}
\label{eq:decomp}
\end{equation}
Here $G^*$ captures candidate-level utility that persists across queries, whereas $H^*$ captures the remaining query-specific compatibility. Thus, two domains can exhibit equally poor Pearson retrieval for fundamentally different reasons: one may contain candidates that are broadly useful regardless of the query, while another may require matching a candidate to the current query. Appendix~\ref{app:proofs} shows that $\E[H^*\mid\mathbf{o}_i]=0$ and, with finite second moments, the two components are orthogonal. This is a population decomposition for interpretation; it does not imply that $G^*$ and $H^*$ are separately identifiable from a finite reranker without additional assumptions.

\subsection{Controls for Relevance Structure}
\paragraph{Candidate Prior.}
For each candidate, Candidate Prior uses its average future compatibility with historical training queries from the same channel (more generally, the same retrieval stratum). It is query-independent and therefore serves as an empirical diagnostic of candidate-global structure. We do not treat it as an estimator of the theoretical $G^*$, and all averages are computed only from temporally admissible historical training data.

\paragraph{Shuffled Future.}
Shuffled Future is paired with the restricted \prstat ablation. It uses the same MLP, observable inputs, Pearson candidate support, optimizer, and model-selection procedure, but applies a nonzero within-channel cyclic shift to the training query futures. The shift preserves the empirical future marginal while breaking the intended query--future correspondence. Under the idealized conditional-independence argument in Appendix~\ref{app:proofs}, the resulting target cannot exploit true query-specific future correspondence.

Accordingly, the \prstat-vs.-Shuffled contrast serves as a controlled diagnostic for query-specific future supervision while holding candidate support and model architecture fixed; it is not interpreted as a direct estimator of $H^*$. Full \prr is excluded from this control because its learned candidate-support expansion would otherwise confound supervision with candidate availability.

\section{Finding 1: Predictive Relevance Improves Pearson Retrieval}
\subsection{Matched Evaluation Protocol}
We evaluate ETTh1, Weather, Electricity, Traffic, Exchange, and Solar at $H\in\{96,192,336,720\}$ with $K=10$ on common frozen chronological manifests (up to 8,192 query--channel pairs per condition). Full \prr uses Top-100 independently from the Pearson and learned branches; the inference-budget control \prbudget reuses the same frozen checkpoints but truncates both branches to Top-50 before unioning, so at most 100 candidates are reranked. The restricted \prstat control instead reranks only the fixed Pearson Top-100 support. All \prr-family comparisons use the same train-normalized series, same-channel memory, stride 8, temporal-admissibility rule, and Top-$K$ evaluation; candidate futures are fully observed before the query.

The primary retrieval metric is \textbf{AnalogFutureMSE}, the mean future distance of the selected Top-$K$ analogs,
\begin{equation}
\operatorname{AnalogFutureMSE}(q)
=
\frac{1}{K}\sum_{i\in\mathcal R_K(q)} d(q,i),
\label{eq:analogfuturemse}
\end{equation}
where $\mathcal R_K(q)$ denotes the retrieved Top-$K$ set. RetrievalForecastMSE after uniformly averaging candidate futures is reported only as a secondary aggregation diagnostic.

\subsection{Retrieval Results and Diagnostics}
Table~\ref{tab:main} gives the central matched comparison. Full \prr improves Pearson in 23/24 conditions (26.68\% mean, 27.11\% median reduction). \prbudget, which uses the same trained \prr model while limiting inference support to at most 100 candidates, retains nearly the same improvement: 23/24 wins with a 26.56\% mean reduction and a mean union size of 94.93. The more restricted \prstat ablation, which reranks only the fixed Pearson Top-100 support, improves Pearson in 20/24 conditions (17.67\% mean). RAFT-ValScale improves Pearson in 15/24, yet full \prr is lower in 22/24. Thus the headline gain is not explained by either a weak Pearson anchor or a larger-than-100 test-time reranking budget.

\begin{table}[t]
\caption{\textbf{Matched retrieval comparison and controls} (AnalogFutureMSE $\downarrow$). Full \prr and PRR-B100 are ours; PRR-B100 is an inference-budget control that caps support at Top-50 $\cup$ Top-50; \prstat reranks Pearson Top-100; RAFT-Val selects its scale on validation. Bold marks the row minimum.}
\label{tab:main}
\centering
\footnotesize
\renewcommand{\arraystretch}{0.82}
\setlength{\tabcolsep}{2.0pt}
\begin{adjustbox}{max width=\linewidth}
\begin{tabular}{llrrrrr}
\toprule
Dataset & $H$ & \textbf{\prr\ (ours)} & \textbf{PRR-B100 (ours)} & \prstat & RAFT-Val & Pearson\\
\midrule
ETTh1 & 96 & \textbf{1.1643} & 1.1699 & 1.1965 & 1.4180 & 1.6123\\
 & 192 & \textbf{1.2229} & 1.2232 & 1.3664 & 1.6023 & 1.8175\\
 & 336 & 1.4816 & \textbf{1.4804} & 1.4861 & 1.7416 & 1.9929\\
 & 720 & \textbf{1.4618} & 1.4677 & 1.7272 & 1.9800 & 2.2541\\
\midrule
Weather & 96 & 0.4316 & \textbf{0.4212} & 0.6707 & 0.6466 & 1.1613\\
 & 192 & \textbf{0.5147} & 0.5333 & 0.7337 & 0.7295 & 1.0628\\
 & 336 & \textbf{0.6748} & 0.6882 & 0.9122 & 0.8307 & 1.2910\\
 & 720 & \textbf{0.8130} & 0.8358 & 1.1676 & 1.0332 & 1.2958\\
\midrule
Electricity & 96 & \textbf{0.7191} & 0.7212 & 0.8751 & 1.1983 & 1.1796\\
 & 192 & \textbf{0.7345} & 0.7359 & 0.8966 & 1.2316 & 1.1957\\
 & 336 & \textbf{0.7648} & 0.7673 & 0.9301 & 1.2775 & 1.2418\\
 & 720 & \textbf{0.8602} & 0.8630 & 0.9769 & 1.3272 & 1.2877\\
\midrule
Traffic & 96 & 1.4696 & \textbf{1.4565} & 1.4755 & 1.9944 & 1.9738\\
 & 192 & 1.4508 & \textbf{1.4263} & 1.5135 & 1.9814 & 2.0045\\
 & 336 & 1.5147 & \textbf{1.5136} & 1.5783 & 2.0729 & 2.0636\\
 & 720 & 1.7351 & \textbf{1.7311} & 1.7571 & 2.1533 & 2.1665\\
\midrule
Exchange & 96 & \textbf{0.1652} & 0.1725 & 0.2206 & 0.2066 & 0.2133\\
 & 192 & 0.4116 & 0.4020 & 0.4276 & \textbf{0.3844} & 0.4234\\
 & 336 & 0.8094 & 0.8107 & 0.7714 & \textbf{0.6725} & 0.7643\\
 & 720 & 1.7770 & \textbf{1.7765} & 2.1310 & 1.8479 & 1.8868\\
\midrule
Solar & 96 & 0.5087 & \textbf{0.4981} & 0.5270 & 0.5955 & 0.5887\\
 & 192 & 0.5406 & \textbf{0.5389} & 0.5411 & 0.5954 & 0.5979\\
 & 336 & 0.5701 & 0.5766 & \textbf{0.5692} & 0.6385 & 0.6238\\
 & 720 & 0.5507 & \textbf{0.5506} & 0.5782 & 0.6614 & 0.6385\\
\bottomrule
\end{tabular}
\end{adjustbox}
\end{table}

Full \prr and \prbudget are nearly indistinguishable on average: \prbudget trails full \prr by only 0.25\% (11 wins, 13 losses), despite reranking fewer than 100 candidates on average. Relative to the restricted \prstat control, \prbudget improves 22/24 conditions (11.32\% mean), indicating that test-time support quality matters more than candidate count alone. Because \prbudget reuses checkpoints trained with full Top-100 $\cup$ Top-100 support, it is an inference-budget control rather than a training-support ablation. The \prstat$\rightarrow$full-\prr increment therefore remains a joint representation-and-support effect (Appendix~\ref{app:prr_decomp}). Matched CRAFT, SARAF, anchored-L2, Candidate Prior, Shuffled-Future, and similarity diagnostics are reported in Appendices~\ref{app:fullresults}--\ref{app:robustness}.

\section{Finding 2: Forecast Utility Depends on Calibrated Trust}
\label{sec:downstream}
\subsection{Validation-Calibrated Integration}
For frozen Direct forecast $\hat{\mathbf y}^D_q$ and the uniform mean $\hat{\mathbf y}^R_q$ of the Top-$K$ \prr-retrieved futures, we use
\begin{equation}
\hat{\mathbf y}^A_q(\beta)=\hat{\mathbf y}^D_q+\beta\left(\hat{\mathbf y}^R_q-\hat{\mathbf y}^D_q\right).
\label{eq:cal_interp}
\end{equation}
For each condition $c=(\text{dataset},H,\text{backbone})$, the coefficient is selected using validation data only,
\begin{equation}
\beta_c^\star=\arg\min_{\beta\in\mathcal B}\operatorname{MSE}_{\mathrm{val},c}\!\left(\hat{\mathbf y}^A(\beta)\right),
\qquad \mathcal B=\{0,0.025,0.05,0.075,0.10,0.15,0.20\}.
\label{eq:trust_cal}
\end{equation}
The bounded grid is fixed before test evaluation. The selected scalar is then frozen for the matched test condition; no test outcome is used for calibration, and no forecaster or retriever is retrained. This scalar is the only downstream quantity adapted per condition: the \prr retriever and all Direct forecasts remain frozen. The value $\beta=0$ is an explicit abstention that returns the Direct forecast exactly. We retain globally fixed $\beta=0.1$ as a controlled comparison that isolates retrieval under common trust.

\subsection{Downstream Results and the Relevance--Utility Gap}
Across all 120 test conditions, calibrated \prr improves Direct in 79, matches Direct in 15, and is worse in 26, with a +1.41\% mean and +0.46\% median MSE gain. Mean gain denotes the macro-average of the per-condition percentage MSE reductions relative to Direct, so each dataset--horizon--backbone condition contributes equally. Thus \prr improves or matches the frozen Direct forecast in 94/120 conditions. Calibration selects exact Direct fallback in 15 conditions and reduces the aggregate number of losses from 41 under fixed trust to 26, while retaining 79 strict wins.

\begin{table}[t]
\caption{\textbf{Validation-calibrated \prr relative to Direct across frozen backbones.} W/T/L counts lower/equal/higher MSE than the same frozen Direct forecast. Positive gain is better.}
\label{tab:backbone_summary}
\centering
\footnotesize
\renewcommand{\arraystretch}{0.90}
\setlength{\tabcolsep}{6.2pt}
\begin{tabular}{lrrr}
\toprule
Backbone & W/T/L & Mean gain & Median gain\\
\midrule
PatchTST & 15/3/6 & -0.07\% & +0.11\%\\
iTransformer & 17/2/5 & +1.78\% & +0.53\%\\
TimeMixer & 16/3/5 & +0.48\% & +0.22\%\\
Seg-MoE & 11/6/7 & +1.56\% & 0.00\%\\
DLinear & 20/1/3 & +3.29\% & +2.18\%\\
\midrule
All & 79/15/26 & +1.41\% & +0.46\%\\
\bottomrule
\end{tabular}
\end{table}

Win count and mean effect size differ: PatchTST wins 15/24 conditions but averages -0.07\% because a few larger losses outweigh its modest gains, whereas DLinear records 20/1/3 W/T/L and +3.29\%. This backbone dependence separates retrieval relevance from downstream complementarity.

The calibrated behavior also varies substantially by dataset. Solar and Weather improve Direct in all 20 conditions, with +5.55\% and +2.39\% mean gains. Electricity improves in 16/20 with 2 ties (+0.40\% mean), and Traffic improves in 11/20 with 6 ties (+0.51\%). Exchange is mixed at 7/6/7 (+0.35\%). ETTh1 is the clearest counterexample: although \prr improves Pearson retrieval at all four horizons in Table~\ref{tab:main}, only 5/20 downstream conditions improve, one abstains, and 14 degrade, for a -0.75\% mean gain. Across horizons, the W/T/L counts are 19/1/10, 21/4/5, 19/7/4, and 20/3/7 for $H=96,192,336,720$, respectively, with positive mean gains at every horizon. Detailed calibrated summaries and the trust-selection distribution are in Appendix~\ref{app:downstream_summaries}. Complete condition-level results for the common fixed-$\beta=0.1$ control are included in the supplementary material.

\begin{figure}[t]
\centering
\includegraphics[width=0.72\linewidth]{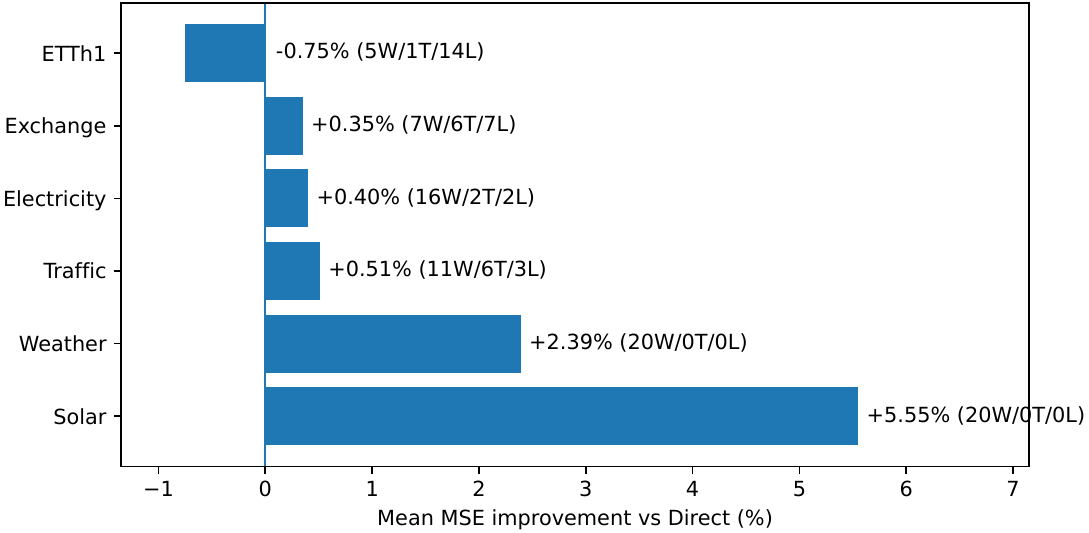}
\caption{\textbf{Dataset-level downstream effect of validation-calibrated \prr relative to Direct.} Bars show mean MSE improvement across five frozen backbones and four horizons; labels report W/T/L counts out of 20. Positive values favor calibrated retrieval augmentation.}
\label{fig:downstream-spectrum}
\end{figure}

The fixed-$\beta=0.1$ control is deliberately retained because it reveals what retrieval does before condition-specific trust is adjusted: it yields the same 79 strict wins but 41 losses and a +0.97\% mean gain. Validation calibration reduces the loss count to 26 and raises the mean gain to +1.41\% while allowing exact Direct fallback in 15 conditions. Validation gain correlates with test gain at $r=0.735$, providing evidence that validation-time utility is informative about test-time retrieval utility. The absence of test-set selection follows from the protocol itself: $\beta$ is selected exclusively on validation data and then frozen before test evaluation.

Large retrieval improvements therefore do not ensure downstream gains for every forecaster. ETTh1 remains negative on average even after calibration, whereas Weather and Solar translate relevance gains into consistent forecast improvements. We therefore separate \emph{which history is relevant} from \emph{how much its retrieved future should influence a particular frozen forecaster}. Appendix~\ref{app:bridge_prr} reports the same-horizon relevance-to-utility bridge under the common fixed-trust control; Appendix~\ref{app:integration_sensitivity} reports calibration diagnostics; and Appendix~\ref{app:mae_sensitivity} gives a secondary loss sensitivity check. Together, these results show that learning which histories are predictively relevant and deciding how strongly those histories should influence a frozen forecaster are distinct problems. Calibration mitigates harmful retrieval use, but it does not eliminate the relevance--utility gap.

\section{Discussion, Limitations, and Scope}

\paragraph{Interpretation and strengths.}
Evaluating PRR as a retrieval-learning framework rather than an end-to-end forecasting architecture separates three often-confounded decisions: which histories are exposed, which are predictively relevant, and how strongly their forecast should influence a direct model. This avoids attributing all gains to one end-to-end retrieval block. Full \prr provides the primary retrieval result. The \prbudget control shows that nearly all of this gain survives with at most 100 test-time candidates; because it reuses the full-support checkpoint, it does not isolate training-time support requirements. The more restricted \prstat ablation reranks only Pearson Top-100, so the \prstat$\rightarrow$\prr comparison jointly adds embedding relations and learned support, while the cardinality audit characterizes candidate availability. Downstream, both retriever and forecasters remain frozen and only validation-selected trust changes, enabling matched comparison across five backbones. ETTh1 is especially informative: retrieval relevance improves substantially while augmentation remains harmful on average, directly separating retrieval quality from forecast utility.

\paragraph{Limitations.}
First, \prr defines privileged relevance through future-trajectory MSE. This choice is aligned with the primary evaluation objective, but another application utility---for example MAE, correlation, event loss, or decision cost---may induce a different relevance ordering. The secondary MAE analysis supports this objective-specific view rather than a metric-independent retrieval benefit (Appendix~\ref{app:mae_sensitivity}). Second, \prr expands the Pearson candidate set but still reranks only the union of Pearson Top-100 and embedding Top-100 histories; it does not jointly search and rerank the complete historical memory. Retrieval is also restricted to the same channel in the aligned benchmark. Third, the predictive encoders and rerankers are trained separately for each dataset and horizon, so the experiments do not establish zero-shot transfer of a universal retriever. Finally, downstream trust is represented by one bounded validation-selected scalar for each dataset--horizon--backbone condition rather than a query-adaptive policy. The upper grid boundary $\beta=0.2$ is selected in 59/120 conditions, suggesting remaining headroom for richer trust models; we deliberately do not enlarge the grid after observing test outcomes. Detailed trust diagnostics are provided in Appendix~\ref{app:integration_sensitivity}.

\paragraph{Scope and implications.}
The frozen-forecaster experiments are a controlled integration study, not a claim of cross-paper forecasting state of the art or universal retrieval superiority. Six datasets, four horizons, and five frozen forecasters use matched chronological manifests and fixed retrieval/calibration protocols. The claim concerns the predictive-relevance principle across these settings, not zero-shot transfer of one universal checkpoint. End-to-end training, full-memory or cross-channel retrieval, utility-specific relevance, and query-adaptive trust remain extensions (Appendix~\ref{app:scope}). Retrieval quality and downstream utility should therefore be reported separately.

\section{Conclusion}

We introduced \textbf{Predictive Relevance Retrieval (PRR)}, a future-supervised, past-only framework for learning which histories should be exposed and ranked. Across six datasets and four horizons, \prr improves Pearson in 23/24 conditions (26.7\% mean AnalogFutureMSE reduction); with five frozen backbones, calibrated trust improves or matches Direct in 94/120 conditions (1.41\% mean relative MSE reduction). Across the evaluated domains, no single past-only retrieval rule dominates, and retrieval and downstream forecasting improvements are empirically distinct objectives.

\section*{Reproducibility Statement}
The main text specifies the Pearson baseline, \prr candidate-support construction, future-compatible target, frozen matched evaluation manifests, and validation-only calibrated downstream protocol. Appendix~\ref{app:data} documents datasets, splits, and temporal admissibility; Appendix~\ref{app:implementation} gives implementation details, including \prr support expansion; Appendices~\ref{app:fullresults}, \ref{app:mechanism}, and~\ref{app:robustness} contain the fixed-support \prstat ablations, controls, strong deterministic similarities, and matched external-retrieval diagnostics underlying Finding~1; and Appendix~\ref{app:strongforecast} contains calibrated aggregate summaries, relevance-to-utility diagnostics, and trust-selection analyses underlying Finding~2. The supplementary material includes the complete condition-level fixed-trust results, frozen manifest audits, the executed PRR-B100, CRAFT, and RAFT retrieval diagnostics, and the notebooks and scripts used to reconstruct the 24-condition retrieval and 120-condition downstream evaluations. Large generated checkpoints, prediction caches, and per-query retrieval arrays are omitted as documented in the supplementary reproducibility instructions. The public reproducibility repository is available at \url{https://github.com/dearyonghoon/which-histories-matter}.

\clearpage
\appendix
\raggedbottom
\section{Proofs and Additional Theory}
\label{app:proofs}
\subsection{Population Characterization of Observable Relevance}
\begin{proposition}[Observable relevance characterization]
Assume $u(q,i)$ has finite second moment. Among all relevance functions measurable from $(\mathbf{o}_q,\mathbf{o}_i)$,
\[
r^*(\mathbf{o}_q,\mathbf{o}_i)=\E[u(q,i)\mid\mathbf{o}_q,\mathbf{o}_i]
\]
uniquely minimizes expected squared relevance-estimation risk up to almost-sure equality.
\end{proposition}
Let $U=u(q,i)$ and $O=(\mathbf{o}_q,\mathbf{o}_i)$. For any square-integrable measurable function $r(O)$, let $r^*(O)=\E[U\mid O]$. Then
\begin{align}
\E[(U-r)^2]
&=\E[(U-r^*+r^*-r)^2]\\
&=\E[(U-r^*)^2]+\E[(r^*-r)^2]
 +2\E[(U-r^*)(r^*-r)].
\end{align}
Because $r^*-r$ is measurable with respect to $O$ and $\E[U-r^*\mid O]=0$,
\begin{equation}
\E[(U-r^*)(r^*-r)]
=\E\left[(r^*-r)\E[U-r^*\mid O]\right]=0.
\end{equation}
Hence
\begin{equation}
\E[(U-r)^2]=\E[(U-r^*)^2]+\E[(r-r^*)^2],
\end{equation}
which is minimized by $r=r^*$ almost surely. For a fixed query and finite candidate set, ranking by $r^*$ therefore ranks candidates by conditional expected future utility. This is a population characterization of the relevance quantity in Eq.~\eqref{eq:bayes}, not a consistency result for the finite-sample learned reranker in Section~3.3.

\subsection{Predictive-Relevance Decomposition}
\begin{proposition}[Predictive relevance decomposition]
For $G^*$ and $H^*$ defined in Eq.~\eqref{eq:ghdef},
\[
\E[H^*(\mathbf{o}_q,\mathbf{o}_i)\mid\mathbf{o}_i]=0.
\]
With finite second moments, $\operatorname{Cov}(G^*,H^*)=0$, and hence
\[
\operatorname{Var}(r^*)=\operatorname{Var}(G^*)+\operatorname{Var}(H^*).
\]
\end{proposition}
By definition,
\begin{equation}
H^*(\mathbf{o}_q,\mathbf{o}_i)=r^*(\mathbf{o}_q,\mathbf{o}_i)-G^*(\mathbf{o}_i),
\end{equation}
with $G^*(\mathbf{o}_i)=\E[U\mid\mathbf{o}_i]$. Applying iterated expectation,
\begin{align}
\E[H^*\mid\mathbf{o}_i]
&=\E[\E[U\mid\mathbf{o}_q,\mathbf{o}_i]\mid\mathbf{o}_i]-\E[U\mid\mathbf{o}_i]\\
&=0.
\end{align}
Since $G^*$ is measurable with respect to $\mathbf{o}_i$,
\begin{equation}
\E[G^*H^*]=\E\{G^*\E[H^*\mid\mathbf{o}_i]\}=0.
\end{equation}
Therefore $G^*$ and $H^*$ are orthogonal in $L^2$. Moreover $\E[H^*]=0$, so the orthogonality implies $\operatorname{Cov}(G^*,H^*)=0$ and the stated variance decomposition. This decomposition is analogous to removing a candidate-observable conditional mean before analyzing the remaining query-dependent interaction; it does not imply that the two terms are separately identifiable from a finite retrieval model without additional assumptions.

\subsection{Ideal Shuffled-Future Argument}
Let $C_q$ denote the semantic stratum (the channel in the controlled benchmarks), and let $\widetilde{\mathbf y}_q$ denote a within-stratum shuffled future. Define the shuffled utility $\widetilde U=-\ell(\widetilde{\mathbf y}_q,\mathbf y_i)$. Under the idealized shuffled-control null, we assume that the shuffled target utility contains no query-specific information beyond candidate-observable information and the stratum:
\begin{equation}
\E[\widetilde U\mid\mathbf{o}_q,\mathbf{o}_i,C_q]
=\E[\widetilde U\mid\mathbf{o}_i,C_q].
\end{equation}
Under this null, shuffled supervision cannot exploit the true query--future correspondence. In the experiment we use a seeded random nonzero cyclic shift within each channel, which preserves each channel's empirical future marginal and guarantees that no query retains its own future when there are at least two queries. The practical control approximates the ideal independence argument; it is not claimed to generate an independent sample for finite time series.

\subsection{Why Retrieval and Aggregation Can Disagree}
For neighbor errors $\mathbf e_j=\mathbf y_{i_j}-\mathbf y_q$, expanding the squared norm of the mean gives
\begin{equation}
\left\|\frac1K\sum_j\mathbf e_j\right\|^2=\frac1{K^2}\left(\sum_j\|\mathbf e_j\|^2+2\sum_{j<k}\mathbf e_j^\top\mathbf e_k\right).
\end{equation}
The first term is proportional to the average individual analog error, while the second depends on pairwise error alignment. Therefore, a retriever can reduce the average individual analog error yet increase the error of the aggregated retrieval forecast by losing beneficial cancellation among neighbors. This motivates AnalogFutureMSE as the primary metric for the relevance-learning question and RetrievalForecastMSE as a downstream diagnostic rather than the optimization target.

\section{Datasets, Splits, and Candidate Construction}
\label{app:data}
\subsection{Benchmark Datasets}
Table~\ref{tab:datasets} summarizes the six generic benchmarks. The aligned long-horizon Pearson/PRR comparison uses all benchmark channels that pass the training-scale validity check and deterministic frozen query--channel manifests, so Electricity and Traffic retain all 321 and 862 channels rather than a reduced subset. Auxiliary mechanism tables are reported under their own stated sampling protocols.

\begin{table}[h]
\caption{Benchmark dataset characteristics and controlled protocol.}
\label{tab:datasets}
\centering
\scriptsize
\begin{tabular}{lrrrrl}
\toprule
Dataset & Rows & Raw ch. & Used ch. & Sampling & Split\\
\midrule
ETTh1 & 17,420 & 7 & 7 & hourly & 12 mo/4 mo/rest\\
Weather & 52,696 & 21 & 21 & 10 min & 70/10/20\\
Electricity & 26,304 & 321 & 321 & hourly & 70/10/20\\
Traffic & 17,544 & 862 & 862 & hourly & 70/10/20\\
Exchange & 7,588 & 8 & 8 & daily & 70/10/20\\
Solar & 52,560 & 137 & 137 & 10 min & 70/10/20\\
\bottomrule
\end{tabular}
\end{table}

For ETTh1, both the aligned long-horizon benchmark and the mechanism study use 8,640 training observations (the first 12 months), 2,880 validation observations (the next 4 months), and all remaining 5,900 observations for test. This is a custom chronological 12-month/4-month/rest split rather than a 12-month/4-month/4-month ETT split. For Weather, the Python-style exclusive split boundaries are \texttt{train\_end=36,887} and \texttt{val\_end=42,157}. The other datasets use chronological 70/10/20 splits. Within the training region, the earliest 60\% forms the initial retrieval memory and the remainder supplies training queries. Validation memory contains the complete training period; test memory contains both the training and validation periods. Every candidate's future end occurs strictly before the relevant query boundary.

\subsection{Window Sampling and Search Caps}
We use $L=96$ throughout the generic retrieval experiments. The main aligned horizons are $H\in\{96,192,336,720\}$; mechanism and robustness diagnostics additionally use the short-horizon $H\in\{24,48,96\}$ suite. For the aligned long-horizon benchmark, Pearson, \prstat, and full \prr all use a common memory-window stride of 8. Auxiliary mechanism experiments retain their separately stated strides and query caps. Retrieval itself uses $L=96$, while frozen evaluation anchors require at least 512 preceding observations so that the same query manifests can be reused across the downstream forecasting study. The aligned comparison uses frozen channel-balanced manifests with up to 4,096 validation and 8,192 test query--channel pairs per dataset--horizon condition. Exchange has fewer admissible pairs at the longest horizons: 3,400 and 328 validation pairs at $H=336$ and $H=720$, respectively, and 6,384 test pairs at $H=720$. Auxiliary pooled-memory diagnostics cap candidate memory at 50,000 windows; the aligned same-channel long-horizon pools are naturally smaller than this cap. Same-channel memory coverage exceeds $M=100$ for every evaluated query channel.

\subsection{Train-Only Normalization and Future Targets}
Each channel is standardized using the mean and standard deviation estimated from the training period. All retained benchmark channels have training-period standard deviation greater than $10^{-6}$; the implementation rejects a benchmark if this validity check is violated. In the final protocol, future trajectories are represented as
\begin{equation}
 y_t(h)=x^{\mathrm{train-norm}}_{t+h}-x^{\mathrm{train-norm}}_t,
 \qquad h=1,\ldots,H.
\end{equation}
We use this fixed train-scale target because it avoids dividing by the standard deviation of an individual past window, which can become arbitrarily small. Appendix~\ref{app:mechanism} documents the Weather scaling diagnostic that motivated this design choice.

\section{Implementation Details}
\label{app:implementation}
\subsection{Observable Statistical Features}
We use seven lightweight past-only statistics as the statistical component of the fusion input. Let the normalized past window be $x_{1:L}$ and let $s$ denote the final quarter of the window. The features are: (1) current value relative to the full-window mean and standard deviation; (2) final-quarter mean minus full-window mean, normalized by the full-window standard deviation; (3) change from the start of the short segment to the current point; (4) full-window endpoint change; (5) ratio of short-segment to full-window first-difference volatility; (6) least-squares slope normalized by window standard deviation; and (7) lag-1 autocorrelation. These statistics are not claimed to be an optimal state representation; they are deliberately lightweight. Full \prr augments their pairwise relations with learned embedding relations and learned candidate-support expansion. The statistics-only reranker described next is used only as the restricted \prstat ablation. Each feature is transformed as $z=(x-\operatorname{median})/\operatorname{IQR}$ using statistics fitted on the training memory. An IQR below $10^{-5}$ is replaced by 1.0, and $z$ is clipped to $[-8,8]$.

\subsection{Fixed-Support PRR-Stat Reranker and Optimization}
For the restricted \prstat ablation, the pair feature has dimension $1+4d_z$ with $d_z=7$. The residual network is a two-layer MLP with hidden dimension 128, LayerNorm, Gaussian error linear unit (GELU) activations, dropout 0.1, and a scalar output. The residual scale is a single global trainable parameter, $\alpha=\operatorname{softplus}(a)$, optimized jointly with the MLP; $a$ is initialized so that $\alpha=0.1$. We train with AdamW using a learning rate of $10^{-3}$, weight decay of $10^{-4}$, batch size 128, a maximum of 30 epochs, patience of 6, and gradient clipping at 5.0. The future target temperature is $\tau_y=0.5$.

Phase A selects the number of epochs using validation AnalogFutureMSE. Phase B reinitializes the \prstat model and refits it from scratch for the selected number of epochs using training and validation retrieval supervision. The test split is evaluated only after this refit. A separate ablation model is trained from scratch for each dataset, horizon, and seed; only architecture and optimization settings are shared. The aligned long-horizon \prstat decomposition uses the frozen seed-0 model for each dataset--horizon condition. The auxiliary short-horizon \prstat and Shuffled suites use five seeds (0--4), while the ETTh1/Weather mechanism study uses three seeds. Seed-level variability for the diagnostic suite is reported below.

\subsection{PRR Candidate Support Expansion}
\label{app:learned_support}
\prr uses a past encoder that takes a single normalized channel window and applies Conv1d blocks $1\!\rightarrow\!64$, $64\!\rightarrow\!128$, and $128\!\rightarrow\!128$ with kernel widths 5, 5, and 3 and stride 2, followed by GELU, adaptive average pooling, a $128\!\rightarrow\!64$ projection, and unit-norm output. Pearson Top-100 and embedding-similarity Top-100 are retrieved independently from the same temporally admissible memory and unioned before final reranking. The fusion feature concatenates the 29-dimensional statistical pair representation with $\mathbf e_q$, $\mathbf e_i$, $\mathbf e_q-\mathbf e_i$, and $|\mathbf e_q-\mathbf e_i|$, giving 285 dimensions. The fusion MLP uses hidden width 384 with LayerNorm, GELU, and dropout 0.1 in two blocks before the scalar residual score. The final Top-10 is selected from the union. All future-derived quantities are restricted to training supervision; validation/test retrieval uses past observations only.

A distinct full-\prr stack is trained for each dataset--horizon condition using train-only histories. The initial 60\% of the training region forms retrieval memory and later train-only anchors are sampled at stride 4. ETTh1 uses all seven channels and all eligible train-only anchors. For Weather, Electricity, Traffic, Exchange, and Solar, the frozen compute-matched training protocol uses at most 64 evenly spaced channels and at most 6,000 channel--anchor pairs; evaluation still uses all channels. The dual encoder sees 100 candidates per query, split evenly between Pearson-hard and randomly sampled admissible histories. Its teacher uses standardized future distances with $\tau_T=0.5$, while the embedding student uses $\tau_S=0.10$. Query and history encoders are optimized jointly for 20 epochs with AdamW (batch 32, learning rate $10^{-3}$, weight decay $10^{-4}$, gradient clipping 5.0).

After encoder training, embeddings are frozen while the fusion reranker is optimized. Train-only query anchors are split chronologically 80/20 for Phase-A model selection. The reranker uses AdamW (batch 64, learning rate $10^{-3}$, weight decay $10^{-4}$), at most 30 epochs, patience 6, and gradient clipping 5.0. Its objective is the listwise term in Eq.~\eqref{eq:loss} plus a Top-$K$-aware pairwise softplus loss with weight 1.0: for each oracle Top-$K$ candidate $i$ (smallest realized future distance) and each remaining union candidate $j$, the pair term averages $\log(1+\exp[-(s_i-s_j)])$ over valid pairs. The positive residual scale is initialized at 0.1 and learned jointly. Finally, the reranker is reinitialized and refit on all train-only supervision records for the selected number of epochs. All full-\prr stacks use seed 0. The supplementary material includes an executable training notebook reproducing this checkpoint schema and records the original ETTh1 $H=192$ checkpoint-reuse provenance.

The support expansion is intentionally modest in architecture but material in search coverage. Each full-\prr branch returns 100 candidates, and the average union contains about 183 distinct histories: approximately 83 are contributed by learned representation retrieval but are absent from Pearson Top-100, with about 17 overlapping histories. To test whether the retrieval gain is merely a consequence of this larger inference pool, \prbudget reuses the same frozen full-\prr checkpoints but truncates each branch to Top-50 before the union. Its union averages 94.93 candidates (median 97, maximum 100), including 44.93 learned-only candidates. Table~\ref{tab:support_cardinality} summarizes both geometries. This is an inference-time budget control; the model is not retrained under the reduced support.

\begin{table}[h]
\caption{Candidate-support geometry. Full \prr uses Top-100 per branch; \prbudget is the inference-only Top-50-per-branch budget control.}
\label{tab:support_cardinality}
\centering
\scriptsize
\begin{tabular}{lrr}
\toprule
Quantity & Full \prr & \prbudget\\
\midrule
Pearson branch & 100 & 50\\
Learned-representation branch & 100 & 50\\
Mean union size & 183.0 & 94.93\\
Mean learned-only candidates & 83.0 & 44.93\\
Mean branch overlap & 17.0 & 5.07\\
Final reranked analogs ($K$) & 10 & 10\\
\bottomrule
\end{tabular}
\end{table}

A separate full-\prr checkpoint is retained for each dataset and horizon. This design isolates the retrieval question from cross-domain representation transfer: the paper claims that future-supervised relevance learning is reproducible across domains, not that one universal encoder transfers zero-shot. At inference, only the query past, historical candidate pasts, Pearson similarity, and learned embeddings are evaluated. Query futures and candidate-future distances enter only through the training target. Because candidate futures are fully historical before the query boundary, the same memory construction can be used to generate supervision without test leakage.

The 285-dimensional fusion input retains the 29-dimensional statistical pair representation used by the restricted \prstat ablation and appends four 64-dimensional representation relations. This makes the comparison interpretable: \prstat tests future-supervised reranking within fixed Pearson support, whereas full \prr adds both representation-level relations and support expansion. Appendix~\ref{app:prr_decomp} reports this decomposition condition by condition.

\subsection{Inference Sequence and Search Complexity}
\label{app:prr_inference}
At validation and test time, PRR follows the same deterministic sequence for every query. First, the query and all admissible same-channel memory windows are represented using train-period normalization. Second, centered normalized past windows produce Pearson scores and the Pearson Top-100 set. Third, the 64-dimensional query embedding is compared with precomputed unit-norm memory embeddings to obtain an independent representation Top-100 set. Fourth, the two lists are unioned and the 285-dimensional pair feature is formed only for distinct candidates in that union. Fifth, the fusion reranker scores the union and returns the final Top-10 analogs. No future value of the test query is touched in any of these steps.

Historical memory embeddings can be precomputed once for each dataset--horizon memory boundary because candidate past windows are fixed. Consequently, the online candidate-generation work consists of two similarity scans over the admissible memory (one scalar Pearson/pattern dot product and one 64-dimensional embedding dot product per candidate), followed by MLP scoring of only about 183 union candidates on average. If $N_m$ is the number of admissible memory windows, $d_e=64$ the embedding dimension, and $|\mathcal U(q)|$ the union size, the query-time search cost is dominated by
\begin{equation}
 O(N_m L)+O(N_m d_e)+O\!\left(|\mathcal U(q)|(d_f h_f+h_f^2)\right),
\end{equation}
where $d_f=285$ is the fusion input dimension and $h_f=384$ its hidden width. The last term is applied to roughly 183 candidates rather than to the full memory. We do not use this expression as a wall-clock claim; it documents which parts of PRR scale with memory size and which remain bounded by the candidate union.

After Top-10 selection, downstream integration requires no additional retrieval training or forecaster update. Candidate relative futures are uniformly averaged to form the historical trajectory. For each dataset--horizon--backbone condition, a scalar trust coefficient is selected on validation from $\{0,0.025,0.05,0.075,0.10,0.15,0.20\}$ and then frozen on test; $\beta=0$ returns Direct exactly. Thus retrieval and forecaster computation remain modular: direct predictions can be cached once, while trust calibration and alternative retrieval variants can be evaluated on the same query manifests without rerunning the forecasting backbone. A globally fixed $0.1$ coefficient is retained as a controlled ablation.

\subsection{Candidate Prior}
For candidate $i$ in channel $c$, Candidate Prior estimates
\begin{equation}
b_i=\E_{q'\in\mathcal Q_{\mathrm{hist},c}} d(q',i),\qquad s_{\mathrm{prior}}(i)=-b_i,
\end{equation}
using historical training-query futures only. For test evaluation, the reference distribution is expanded to include the training and validation data after model selection, consistent with the available historical information. Candidate futures are historical and fully observed before test; no test query future is used to calculate the prior.

\subsection{Shuffled-Future Control}
Within each semantic channel, a seeded random nonzero cyclic shift is applied to the query futures. The same candidate pools and inputs are retained. Phase A and Phase B use independently seeded nonzero shifts within their respective historical supervision partitions. Test futures are never shuffled because they are used only for evaluation.

\subsection{Ranking Metrics and Statistical Testing}
For the fixed-support \prstat diagnostics, realized future distances within the Pearson Top-$M$ candidate set are standardized across candidates as $\widetilde d_{qi}$, and ranking gain is defined as $g_{qi}=\exp(-\widetilde d_{qi}/\tau_y)$. Normalized discounted cumulative gain (NDCG)@$K$ computes discounted cumulative gain for the retrieved Top-$K$ using $g_{qi}$ and normalizes it by the ideal Top-$K$ ordering under the same gains. OracleRecall@$K$ is the fraction of retrieved Top-$K$ candidates that overlap the oracle Top-$K$, where the oracle consists of the $K$ candidates in $\mathcal C_M(q)$ with smallest realized future distance. These ranking metrics are used only for the fixed-support diagnostic experiments; the headline full-\prr comparison uses AnalogFutureMSE on the final retrieved Top-$K$.

For the final 24-condition aligned retrieval comparison, all methods reuse the exact same frozen query--channel evaluation manifests; we emphasize effect sizes, win counts, and sign consistency. No new significance symbols are attached to Table~\ref{tab:main}. The auxiliary short-horizon mechanism and SARAF-Matched diagnostics use their stated moving-block intervals within their own matched protocols.

\subsection{Frozen Evaluation Manifest and Reproducibility Audit}
\label{app:eval_manifest}
The aligned long-horizon evaluation is driven by deterministic query--channel manifests rather than by independently resampling queries for different methods. All aligned retrieval methods use the common memory-window stride of 8 specified in Appendix~\ref{app:data}. Table~\ref{tab:manifest_audit} records the effective test coverage. Each manifest stores the query anchor and channel index; full \prr, \prbudget, \prstat, Pearson, the matched external-retrieval diagnostics, and all downstream Direct/PRR comparisons reuse these exact evaluation pairs. This prevents small sampling changes from being mistaken for retrieval improvements. Exchange at $H=720$ is the only test condition with fewer than 8,192 pairs because the admissible chronological test range is shorter.

\begin{table}[h]
\caption{Frozen test-manifest audit for the aligned long-horizon benchmark. Entries under each horizon are the number of matched query--channel pairs.}
\label{tab:manifest_audit}
\centering
\scriptsize
\setlength{\tabcolsep}{5.5pt}
\begin{tabular}{lrrrrr}
\toprule
Dataset & Channels & $H=96$ & $H=192$ & $H=336$ & $H=720$\\
\midrule
ETTh1 & 7 & 8192 & 8192 & 8192 & 8192\\
Weather & 21 & 8192 & 8192 & 8192 & 8192\\
Electricity & 321 & 8192 & 8192 & 8192 & 8192\\
Traffic & 862 & 8192 & 8192 & 8192 & 8192\\
Exchange & 8 & 8192 & 8192 & 8192 & 6384\\
Solar & 137 & 8192 & 8192 & 8192 & 8192\\
\bottomrule
\end{tabular}
\end{table}

The audit also reproduces the chronological normalization and memory boundaries before scoring. For test queries, retrieval memory is restricted to train+validation history and every candidate satisfies ``candidate anchor + horizon $\leq$ memory boundary.'' Pearson Top-10 is then computed from the full admissible same-channel memory, whereas \prr forms the union of Pearson Top-100 and learned-representation Top-100 before reranking. No forecaster inference is required to reproduce the retrieval comparison itself.

\section{Retrieval Ablations and External-Baseline Diagnostics}
\label{app:fullresults}
This appendix separates the main full-\prr result from fixed-support mechanism diagnostics. We first summarize the long-horizon controls already reported condition by condition in Table~\ref{tab:main}, then return to the short-horizon $H\in\{24,48,96\}$ suite used to isolate future supervision and alternative similarity rules.

\subsection{Long-Horizon Fixed-Support and Candidate-Budget Controls}
\label{app:prr_decomp}
Table~\ref{tab:main} already reports the condition-level long-horizon comparison. Full \prr provides the primary result. The \prbudget variant uses the same trained \prr model while capping test-time support at 100 candidates and retains nearly the same retrieval gain, showing that the headline improvement is not explained simply by a larger inference candidate count. The more restricted \prstat control keeps exactly the Pearson Top-100 support and isolates future-supervised reranking within that fixed support. \prbudget improves \prstat in 22/24 conditions, while full \prr improves \prstat in 22/24; because the latter comparison jointly adds embedding relations and learned candidate support, it should not be interpreted as a pure support-expansion effect. The support geometry underlying these controls is reported in Appendix~\ref{app:implementation}.

\subsection{Four-Dataset Confirmatory Fixed-Support Suite}
\label{app:fixed_support_suite}
Electricity, Traffic, Exchange, and Solar form the common five-seed confirmatory suite for fixed-support diagnostics. ETTh1 and Weather are evaluated separately in Section~D.5 under the dedicated mechanism protocol, where they provide deliberately contrasting query-specific and candidate-global regimes. Thus all six benchmark datasets are covered by the short-horizon diagnostics, but the two mechanism datasets are reported separately to preserve the distinction between confirmatory and mechanism-oriented analyses.

Table~\ref{tab:full_confirm} reports NDCG@$K$, OracleRecall@$K$, and uniform-neighbor RetrievalForecastMSE for the four-dataset suite. Exchange illustrates strong candidate-global structure: the diagnostic Candidate Prior achieves lower AnalogFutureMSE than Pearson and \prstat despite ignoring the current query. Solar shows the opposite regime, where Candidate Prior is poor and \prstat attains substantially lower AnalogFutureMSE together with high NDCG. Candidate Prior is used only as a mechanism diagnostic because it relies on historical future information.

\begin{table}[h]
\caption{Four-dataset short-horizon fixed-support diagnostics. For \prstat and Shuffled, values are averaged over five seeds.}
\label{tab:full_confirm}
\centering
\scriptsize
\begin{adjustbox}{max width=\linewidth}
\begin{tabular}{llrrrrrrrr}
\toprule
Data & $H$ & Pear. A & Stat A & Shuf A & Prior A & Pear. F & Stat F & Stat NDCG & Stat Recall\\
\midrule
Electricity&24&0.4138&0.3360&0.4381&0.3469&0.2398&0.1961&0.5132&0.2106\\
&48&0.4557&0.3753&0.5025&0.3828&0.2648&0.2197&0.5096&0.2127\\
&96&0.4771&0.3924&0.5545&0.4118&0.2725&0.2263&0.5040&0.2192\\
Traffic&24&0.8557&0.8163&1.1204&0.9088&0.5895&0.5851&0.6072&0.2407\\
&48&0.8691&0.8609&1.0889&0.9157&0.6040&0.6090&0.5451&0.2490\\
&96&0.8751&0.8634&1.0919&0.9057&0.6115&0.6185&0.4943&0.2598\\
Exchange&24&0.0587&0.0466&0.0443&0.0249&0.0262&0.0253&0.5801&0.1082\\
&48&0.1110&0.0930&0.0894&0.0484&0.0494&0.0472&0.5629&0.1136\\
&96&0.2120&0.2012&0.1999&0.0989&0.0991&0.0977&0.5389&0.1105\\
Solar&24&0.3050&0.2034&0.8833&0.4081&0.1806&0.1268&0.7922&0.3101\\
&48&0.5253&0.3529&1.1849&0.8337&0.3330&0.2246&0.7868&0.3068\\
&96&0.6662&0.4713&1.3160&1.0164&0.4344&0.2898&0.6838&0.2970\\
\bottomrule
\end{tabular}
\end{adjustbox}
\vspace{1mm}
\footnotesize Pear./Stat/Shuf/Prior A denote AnalogFutureMSE; Pear./Stat F denote RetrievalForecastMSE.
\end{table}

\subsection{External Retrieval Baseline: SARAF-Matched}
\label{app:saraf}
We apply the public SARAF retrieval rule to the same frozen Pearson Top-100 pools, queries, same-channel admissibility, fixed train-scale target, $M=100$, and $K=10$. We restrict this comparison to the four-dataset five-seed confirmatory suite; ETTh1 and Weather are reserved for the separate mechanism study in Section~D.5. Its MMR coefficient is $\lambda=0.3+0.6s$, where the dataset-level $(s,\lambda)$ values are (0.6504,0.6902) for Electricity, (0.6053,0.6632) for Traffic, (0.4309,0.5585) for Exchange, and (0.3567,0.5140) for Solar. Results average five stochastic MMR seeds.

\prstat has lower AnalogFutureMSE in all 12 tasks, and every condition-wise 95\% moving-block interval for $\mathrm{MSE}_{\mathrm{SARAF}}-\mathrm{MSE}_{\mathrm{PR\text{-}Stat}}$ is positive. These intervals are condition-wise diagnostics rather than simultaneous family-wise confidence intervals. SARAF-Matched improves Pearson only for Exchange at $H=24$ (1.06\%, not significant). This is a matched evaluation of SARAF's retrieval rule, not a reproduction of its forecasting benchmark.

\begin{table}[h]
\caption{External retrieval comparison under the matched protocol (AnalogFutureMSE $\downarrow$). The confidence interval (CI) is the condition-wise 95\% moving-block bootstrap interval for SARAF-Matched minus \prstat.}
\label{tab:saraf_full}
\centering
\scriptsize
\setlength{\tabcolsep}{3.0pt}
\begin{adjustbox}{max width=\linewidth}
\begin{tabular}{llrrrrr}
\toprule
Dataset & $H$ & Pearson & SARAF-M & \prstat & \multicolumn{1}{c}{\shortstack{PRR-Stat gain\\vs. SARAF}} & \multicolumn{1}{c}{95\% CI}\\
\midrule
Electricity & 24 & 0.4138 & 0.4807 & \textbf{0.3360} & 30.1\% & [0.1359,0.1518]\\
            & 48 & 0.4557 & 0.5322 & \textbf{0.3753} & 29.5\% & [0.1472,0.1649]\\
            & 96 & 0.4771 & 0.5522 & \textbf{0.3924} & 29.0\% & [0.1484,0.1665]\\
Traffic     & 24 & 0.8557 & 1.0578 & \textbf{0.8163} & 22.8\% & [0.2239,0.2627]\\
            & 48 & 0.8691 & 1.0710 & \textbf{0.8609} & 19.6\% & [0.1959,0.2306]\\
            & 96 & 0.8751 & 1.0444 & \textbf{0.8634} & 17.3\% & [0.1671,0.2008]\\
Exchange    & 24 & 0.0587 & 0.0581 & \textbf{0.0466} & 19.7\% & [0.0096,0.0127]\\
            & 48 & 0.1110 & 0.1132 & \textbf{0.0930} & 17.8\% & [0.0167,0.0230]\\
            & 96 & 0.2120 & 0.2185 & \textbf{0.2012} & 7.9\% & [0.0086,0.0265]\\
Solar       & 24 & 0.3050 & 0.4769 & \textbf{0.2034} & 57.3\% & [0.2540,0.2908]\\
            & 48 & 0.5253 & 0.9385 & \textbf{0.3529} & 62.4\% & [0.5508,0.6063]\\
            & 96 & 0.6662 & 1.0954 & \textbf{0.4713} & 57.0\% & [0.5900,0.6478]\\
\bottomrule
\end{tabular}
\end{adjustbox}
\end{table}

\subsection{Seed Stability}
\prstat exhibits low five-seed variability across the four-dataset short-horizon suite, with all coefficients of variation below 3\%. Shuffled is more variable in several settings, particularly Electricity. Complete seed-level means and coefficients of variation are reported in Appendix~\ref{app:seed_stability} and Table~\ref{tab:seed_full}.

\subsection{ETTh1 and Weather Mechanism Controls}
ETTh1 and Weather are not omitted from the short-horizon analysis; they are separated because their experiments were designed specifically to diagnose the relevance mechanism. Table~\ref{tab:ett_weather} reports the short-horizon same-channel fixed-scale study. On ETTh1, \prstat beats Shuffled by 26.2--30.2\% while Candidate Prior is worse than Pearson. Weather shows the opposite: Prior improves Pearson by 65.3--72.1\%, and Shuffled is lower than \prstat at all three horizons. This contrast motivates the candidate-global/query-specific interpretation.

\begin{table}[h]
\caption{Same-channel fixed-scale mechanism controls for ETTh1 and Weather (AnalogFutureMSE $\downarrow$). \prstat and Shuffled are averaged over three seeds; Pearson and Candidate Prior are deterministic diagnostics.}
\label{tab:ett_weather}
\centering
\scriptsize
\begin{tabular}{llrrrr}
\toprule
Dataset & $H$ & Pearson & Prior & \prstat & Shuffled\\
\midrule
ETTh1&24&0.9308&1.1888&\textbf{0.7820}&1.0593\\
&48&1.0527&1.1882&\textbf{0.8679}&1.2434\\
&96&1.1959&1.2183&\textbf{0.9954}&1.3874\\
\midrule
Weather&24&0.3963&\textbf{0.1375}&0.3565&0.3150\\
&48&0.7132&\textbf{0.1988}&0.5772&0.5607\\
&96&0.9131&\textbf{0.2579}&0.8063&0.6277\\
\bottomrule
\end{tabular}
\end{table}
\FloatBarrier

\section{Mechanism Analyses and Negative Results}
\label{app:mechanism}
This section records development-stage diagnostics used to understand target scaling and fixed-support relevance mechanisms. These experiments are not additional full-\prr claims; the headline benchmark uses the frozen protocol described in Section~3 and Appendix~B.

\subsection{Weather Reveals a Small-Denominator Pathology in the Local-Scale Target}
During exploratory development, the local-scale target $(x_{t+h}-x_t)/\sigma_{\mathrm{past}}$ was found to be numerically unstable on low-variance Weather windows. Because its denominator is window specific, near-constant pasts can produce extreme target magnitudes. A robustness experiment that imposed a train-memory quantile floor substantially reduced this numerical tail, but \prstat still did not consistently beat Pearson or Shuffled on Weather.

The final protocol instead uses the fixed train-channel scale, which depends only on training-period normalization and removes the per-window denominator. All headline long-horizon comparisons were subsequently evaluated under this single frozen target definition. Under the final same-channel fixed-scale target, Candidate Prior remains the strongest diagnostic on Weather (Table~\ref{tab:ett_weather}); Appendix~J.1 records the retrospective scaling analysis in detail.

\subsection{Same-Channel Restriction Does Not Fully Explain Weather}
Under the fixed target, same-channel retrieval outperforms cross-channel retrieval on Weather at $H=24$ and $96$, but not at $H=48$. Thus cross-channel mismatch is not a sufficient explanation of the Weather behavior, and same-channel retrieval is used as a controlled semantic setting rather than claimed as a universally superior candidate policy.

\subsection{Candidate-Prior Deconfounding as an Identification Diagnostic}
Subtracting an empirical candidate prior before training preserves an approximately 21\% ETTh1 \prstat-over-Shuffled advantage across the three short horizons, supporting residual query-specific signal under this diagnostic. However, deconfounding worsens absolute AnalogFutureMSE on both ETTh1 and Weather, and on Weather it does not restore an advantage over the corresponding shuffled control. Prior subtraction is therefore used only as a mechanism diagnostic, not as the retrieval objective.

The full horizon-wise values are reported in Appendix~J.2. Thus Eq.~\ref{eq:decomp} is a structural view, not a prescription for a $G+H$ architecture.

\section{Frozen-Forecaster Integration Details}
\label{app:strongforecast}
\subsection{Validation-Calibrated Main Protocol}
\label{app:calibrated_protocol}
For every backbone, the retrieval forecast is the uniform mean of the Top-10 \prr historical futures. If candidate $i$ has endpoint-anchored relative future $\mathbf y_i$, the historical-memory forecast is $\hat{\mathbf y}^{R}_q=x_q(t)+K^{-1}\sum_{i\in\mathcal N_K(q)}\mathbf y_i$ on the normalized forecast scale. The main downstream protocol selects one scalar $\beta_c^\star$ per dataset--horizon--backbone condition using Eq.~\ref{eq:trust_cal} on validation only. Direct forecasts are computed once and frozen; \prr is integrated on exactly the same query--channel pairs. No test outcome is used to choose $\beta$.

The evaluation covers PatchTST, iTransformer, TimeMixer, Seg-MoE, and DLinear on ETTh1, Weather, Electricity, Traffic, Exchange, and Solar for $H\in\{96,192,336,720\}$. Direct cache coverage is 120/120 conditions. The final comparison performs no forecaster retraining, repeated inference, or retriever refitting. Calibration is a seven-point scalar sweep over cached validation predictions.

Table~\ref{tab:trust_distribution} records the selected trust distribution. Fifteen conditions choose $\beta=0$ and therefore abstain to Direct; 59 choose the upper grid value $0.2$. We report this saturation explicitly and do not enlarge the grid after inspecting test results.

\begin{table}[h]
\caption{Validation-selected trust coefficients for the 120 dataset--horizon--backbone conditions. Selection uses validation MSE only.}
\label{tab:trust_distribution}
\centering
\scriptsize
\begin{tabular}{lrrrrrrr}
\toprule
$\beta$ & 0 & .025 & .05 & .075 & .10 & .15 & .20\\
\midrule
Conditions & 15 & 12 & 12 & 6 & 5 & 11 & 59\\
\bottomrule
\end{tabular}
\end{table}

A common-trust control is retained to separate relevance from calibration. With fixed $\beta=0.1$, \prr gives 79W/0T/41L, +0.97\% mean gain, and +0.79\% median gain. Condition calibration gives 79W/15T/26L, +1.41\% mean gain, and +0.46\% median gain. The number of strict wins is unchanged. Fifteen conditions select $\beta=0$ and therefore exactly recover Direct, while the aggregate loss count decreases from 41 to 26.

\subsection{Dataset, Backbone, and Horizon Summaries}
\label{app:downstream_summaries}
Table~\ref{tab:downstream_dataset_new} reports calibrated \prr relative to Direct by dataset. Weather and Solar improve in every condition; Electricity and Traffic become positive on average after calibration, while ETTh1 remains the principal failure case.

\begin{table}[h]
\caption{Dataset-level downstream summary for validation-calibrated \prr relative to Direct. Positive mean/median gain is better.}
\label{tab:downstream_dataset_new}
\centering
\scriptsize
\begin{tabular}{lrrr}
\toprule
Dataset & W/T/L & Mean gain & Median gain\\
\midrule
ETTh1 & 5/1/14 & -0.75\% & -0.43\%\\
Weather & 20/0/0 & +2.39\% & +1.49\%\\
Electricity & 16/2/2 & +0.40\% & +0.29\%\\
Traffic & 11/6/3 & +0.51\% & +0.08\%\\
Exchange & 7/6/7 & +0.35\% & 0.00\%\\
Solar & 20/0/0 & +5.55\% & +5.27\%\\
\bottomrule
\end{tabular}
\end{table}

\begin{table}[h]
\caption{Forecast-horizon downstream summary for validation-calibrated \prr relative to Direct.}
\label{tab:downstream_horizon_new}
\centering
\scriptsize
\begin{tabular}{lrrr}
\toprule
$H$ & W/T/L & Mean gain & Median gain\\
\midrule
96 & 19/1/10 & +1.14\% & +0.25\%\\
192 & 21/4/5 & +1.22\% & +0.41\%\\
336 & 19/7/4 & +1.59\% & +0.43\%\\
720 & 20/3/7 & +1.68\% & +0.89\%\\
\bottomrule
\end{tabular}
\end{table}

Table~\ref{tab:downstream_backbone_new} gives the complementary backbone view. DLinear and iTransformer have the largest average gains, whereas PatchTST is approximately neutral on average despite 15 strict wins and 3 abstentions.

\begin{table}[h]
\caption{Backbone-level downstream summary for validation-calibrated \prr relative to Direct across six datasets and four horizons (24 conditions per backbone).}
\label{tab:downstream_backbone_new}
\centering
\scriptsize
\begin{tabular}{lrrr}
\toprule
Backbone & W/T/L & Mean gain & Median gain\\
\midrule
PatchTST & 15/3/6 & -0.07\% & +0.11\%\\
iTransformer & 17/2/5 & +1.78\% & +0.53\%\\
TimeMixer & 16/3/5 & +0.48\% & +0.22\%\\
Seg-MoE & 11/6/7 & +1.56\% & 0.00\%\\
DLinear & 20/1/3 & +3.29\% & +2.18\%\\
\bottomrule
\end{tabular}
\end{table}

\paragraph{Fixed-trust interaction control.}
For comparison, Table~\ref{tab:downstream_matrix_new} retains the dataset--backbone interaction under the common $\beta=0.1$ control. This table is diagnostic rather than the calibrated headline result: it shows where the same amount of retrieved-history trust helps or hurts before validation calibration is allowed.

\begin{table}[h]
\caption{Mean downstream MSE gain (\%) under the fixed $\beta=0.1$ control, averaged over four horizons. Positive values favor \prr.}
\label{tab:downstream_matrix_new}
\centering
\scriptsize
\setlength{\tabcolsep}{4.1pt}
\begin{tabular}{lrrrrr}
\toprule
Dataset & PatchTST & iTransformer & TimeMixer & Seg-MoE & DLinear\\
\midrule
ETTh1 & -0.59\% & +0.73\% & +0.66\% & +0.53\% & -0.09\%\\
Weather & +1.28\% & +2.78\% & +1.32\% & +1.13\% & +2.49\%\\
Electricity & -1.39\% & -0.03\% & -0.82\% & -2.20\% & +0.76\%\\
Traffic & -0.49\% & -0.17\% & -0.56\% & -1.89\% & +2.14\%\\
Exchange & -0.05\% & +0.29\% & +0.87\% & -0.66\% & +4.45\%\\
Solar & +2.10\% & +4.55\% & +1.39\% & +7.07\% & +3.52\%\\
\bottomrule
\end{tabular}
\end{table}

\subsection{Same-Horizon Relevance-to-Utility Bridge}
\label{app:bridge_prr}
Table~\ref{tab:bridge_prr} aligns Pearson$\rightarrow$\prr retrieval gain with Direct$\rightarrow$\prr downstream gain under the common fixed-$\beta=0.1$ control. Holding trust fixed is useful diagnostically because it exposes the relevance--utility gap before calibration. The downstream number is the mean percentage MSE gain across five frozen backbones, and the final column counts improved backbones.

\begin{table}[H]
\caption{Fixed-trust same-horizon bridge between retrieval relevance and downstream utility. Retrieval gain is the AnalogFutureMSE reduction of \prr relative to Pearson. Downstream gain uses the common $\beta=0.1$ control and is averaged across five frozen backbones; Wins counts improved backbones out of five.}
\label{tab:bridge_prr}
\centering
\scriptsize
\setlength{\tabcolsep}{5.0pt}
\begin{tabular}{llrrr}
\toprule
Dataset & $H$ & Retrieval gain & Downstream gain & Wins\\
\midrule
ETTh1 & 96 & +27.78\% & -0.08\% & 2/5\\
  & 192 & +32.72\% & +0.21\% & 4/5\\
  & 336 & +25.66\% & +0.19\% & 4/5\\
  & 720 & +35.15\% & +0.67\% & 3/5\\
\midrule
Weather & 96 & +62.84\% & +2.08\% & 5/5\\
  & 192 & +51.57\% & +1.93\% & 5/5\\
  & 336 & +47.73\% & +1.63\% & 5/5\\
  & 720 & +37.26\% & +1.57\% & 5/5\\
\midrule
Electricity & 96 & +39.04\% & -1.00\% & 1/5\\
  & 192 & +38.57\% & -1.23\% & 1/5\\
  & 336 & +38.41\% & -0.66\% & 1/5\\
  & 720 & +33.20\% & -0.05\% & 2/5\\
\midrule
Traffic & 96 & +25.54\% & -0.46\% & 1/5\\
  & 192 & +27.62\% & +0.21\% & 3/5\\
  & 336 & +26.60\% & -0.06\% & 2/5\\
  & 720 & +19.91\% & -0.47\% & 1/5\\
\midrule
Exchange & 96 & +22.54\% & +0.69\% & 3/5\\
  & 192 & +2.79\% & +0.61\% & 4/5\\
  & 336 & -5.90\% & +2.04\% & 5/5\\
  & 720 & +5.82\% & +0.59\% & 2/5\\
\midrule
Solar & 96 & +13.59\% & +2.59\% & 5/5\\
  & 192 & +9.58\% & +3.46\% & 5/5\\
  & 336 & +8.61\% & +4.00\% & 5/5\\
  & 720 & +13.76\% & +4.85\% & 5/5\\
\bottomrule
\end{tabular}
\end{table}

Two contrasts are especially diagnostic. Electricity and Traffic show substantial retrieval gains (roughly 20--39\%) but translate them into little or negative mean downstream gain when every condition is forced to use the same $\beta=0.1$. Weather shows the largest retrieval gains and improves downstream forecasting throughout, while Solar shows smaller relevance gains but consistent forecast improvements. These patterns motivate the calibrated main protocol: validation can reduce or abstain from retrieval trust when the fixed bridge is harmful, but it does not erase all failures (ETTh1 remains negative on average).

\subsection{Secondary MAE Sensitivity}
\label{app:mae_sensitivity}
The retrieval target and trust calibration are both selected using squared trajectory error, so MSE is the metric aligned with the main protocol. As a secondary robustness check, the same fixed-$\beta=0.1$ predictions improve MAE in 56/120 conditions and degrade 64/120, with a near-neutral -0.19\% mean change. This is consistent with the utility-specific interpretation: relevance and trust optimized for squared error need not improve absolute error. Complete condition-level MSE and MAE results are included in the supplementary material. We leave utility-specific retraining for alternative losses to future work rather than selecting a metric after observing test outcomes.

\subsection{Integration Sensitivity}
\label{app:integration_sensitivity}
The calibrated protocol was chosen through a cache-only validation study, not by test optimization. The seven-point grid is fixed before test evaluation, and each of the 120 condition-specific coefficients is selected by validation MSE. Across the 120 conditions, validation and test percentage gains have Pearson correlation $r=0.735$, and their improvement signs agree in 94/120 conditions. Table~\ref{tab:integration_compare} compares the calibrated result with the common-trust control.

\begin{table}[h]
\caption{Trust-integration comparison on the same 120 matched test conditions. Fixed-$0.1$ uses one common coefficient; Condition-Cal selects one bounded coefficient per dataset--horizon--backbone condition using validation only.}
\label{tab:integration_compare}
\centering
\scriptsize
\begin{tabular}{lrrrr}
\toprule
Protocol & W/T/L & Mean gain & Median gain & Non-degraded\\
\midrule
Fixed $\beta=0.1$ & 79/0/41 & +0.97\% & +0.79\% & 79/120\\
Condition-Cal & 79/15/26 & +1.41\% & +0.46\% & 94/120\\
\bottomrule
\end{tabular}
\end{table}

The calibrated result should be interpreted as bounded trust selection, not a learned gating network. Its main effect is defensive: the strict-win count remains 79, the loss count decreases from 41 to 26, and 15 conditions select $\beta=0$ for an exact Direct fallback. An earlier unconstrained closed-form scalar diagnostic could select very large coefficients and fail sharply under distribution shift; bounding the validation grid avoids that instability. Because 59/120 selected coefficients equal the upper bound $0.2$, a wider range could in principle change the result, but we intentionally do not enlarge the grid after observing test behavior.

\subsection{Backbone Provenance and Recipe Audit}
\label{app:backbone_provenance}
The downstream study is designed as a paired frozen-forecaster integration benchmark rather than as an absolute forecasting leaderboard. PatchTST, iTransformer, TimeMixer, Seg-MoE, and DLinear are treated as frozen direct forecasters. Retrieval results never choose direct-model hyperparameters, epochs, or checkpoints. PatchTST uses a patch-based channel-independent Transformer; iTransformer models variates as tokens; TimeMixer uses decomposable multiscale mixing; Seg-MoE uses segment-wise sparse mixture-of-experts routing; and DLinear provides a lightweight decomposition-based linear reference. The matched Direct/PRR comparison reuses the same saved direct predictions for both sides of every pair.

Seg-MoE requires additional recipe bookkeeping because several public configurations are dataset-specific. Table~\ref{tab:segmoe_recipes_prr} records the frozen provenance used for the aligned benchmark. Electricity uses the public ECL-base recipe, Weather the public Weather-small recipe, Traffic its published dataset-specific base recipe, and ETTh1 the small configuration. No public Solar or Exchange recipe was available in the experimental source used here, so their surrogates were fixed before target-test inspection: Solar reuses ECL-base and Exchange reuses ETTh1-small. No target-test hyperparameter search is performed when attaching \prr. Across these runs, block size is 512, patch width is 8, and width factor 4 yields 32-step autoregressive generation. Backend-safety changes (non-fused AdamW, math-only SDPA, and BF16 only after a numerical smoke test) do not alter the declared model recipe.

\begin{table}[h]
\caption{Seg-MoE recipe provenance for the six aligned downstream datasets. Solar and Exchange use predeclared surrogates because no public target-dataset recipe was available; neither is tuned on target-test results.}
\label{tab:segmoe_recipes_prr}
\centering
\scriptsize
\setlength{\tabcolsep}{3.0pt}
\begin{adjustbox}{max width=\linewidth}
\begin{tabular}{lllcrrc}
\toprule
Dataset & Recipe source & Model & Segment sizes & Epochs & LR (max/min) & Batch\\
\midrule
Electricity & public ECL-base & base & [5,5,4,4,3,3] & 15 & $2.6\times10^{-5}/3.2\times10^{-6}$ & 14\\
Traffic & published Traffic recipe & base & [5,5,4,3,2,2] & 10 & $3.2\times10^{-5}/1.2\times10^{-6}$ & 8$^\dagger$\\
Weather & public Weather-small & small & [3,5,5,5] & 20 & $3.2\times10^{-4}/1.2\times10^{-5}$ & 256\\
ETTh1 & Table 8 + SmallConfig & small & [4,5,5,4] & 20 & $3.2\times10^{-4}/1.2\times10^{-4}$ & 256\\
Solar & ECL-base surrogate$^*$ & base & [5,5,4,4,3,3] & 15 & $2.6\times10^{-5}/3.2\times10^{-6}$ & 14\\
Exchange & ETTh1-small surrogate$^*$ & small & [4,5,5,4] & 20 & $3.2\times10^{-4}/1.2\times10^{-4}$ & 256\\
\bottomrule
\end{tabular}
\end{adjustbox}
\vspace{1mm}
\begin{minipage}{0.96\linewidth}
\footnotesize $^*$ Frozen before target-test evaluation; no target-test tuning. $^\dagger$ Traffic uses micro-batch 4 $\times$ accumulation 2, preserving effective batch 8.
\end{minipage}
\end{table}

For DLinear, the direct model uses the official moving-average decomposition and the frozen chronological training recipe; retrieval does not alter its fit. The same principle applies to the neural backbones: PRR is attached only after direct predictions are frozen. This separation ensures that every Direct/\prr pair differs only through the addition of the historical-memory signal rather than through direct-forecaster retuning.

\section{Additional Ablations and Similarity Robustness}
\label{app:robustness}
Unless otherwise stated, this section uses the short-horizon $H\in\{24,48,96\}$ protocol to diagnose architecture, supervision, similarity, and candidate-pool effects with the restricted \prstat ablation. These short-horizon significance statements are condition-wise moving-block diagnostics rather than simultaneous family-wise claims. The aligned long-horizon Pearson-vs.-\prr comparison is reported in Table~\ref{tab:main}; the stronger anchored-L2 stress tests below use the same frozen long-horizon manifests.

\subsection{Fixed-Support Architecture and Supervision Ablation}
Table~\ref{tab:app_ablation} reuses the four-dataset five-seed confirmatory suite of Appendix~\ref{app:fixed_support_suite} to separate observable context from correct future supervision. Pearson+Context uses the same seven statistical features but no MLP and no future labels: Pearson and context-distance ranks are fused with equal weight. Shuffled-Future and \prstat use the same MLP, features, candidate pool, and optimization; only the query--future correspondence differs. Pearson+Context improves Pearson on all 12 tasks, with condition-wise significant gains on eight; \prstat improves Shuffled-Future on 9/12, with all nine gains condition-wise significant; and \prstat improves Pearson+Context on 11/12, with eight condition-wise significant gains.

\begin{table}[h]
\caption{Fixed-support architecture/supervision ablation (AnalogFutureMSE $\downarrow$). \prstat and Shuffled-Future are five-seed means; Pearson and Pearson+Context are deterministic diagnostics.}
\label{tab:app_ablation}
\centering
\scriptsize
\begin{adjustbox}{max width=\linewidth}
\begin{tabular}{llrrrr}
\toprule
Dataset & $H$ & Pearson & Pearson+Context & Shuffled-Future & \prstat\\
\midrule
Electricity&24&0.4138&0.3859&0.4381&\textbf{0.3360}\\
&48&0.4557&0.4220&0.5025&\textbf{0.3753}\\
&96&0.4771&0.4370&0.5545&\textbf{0.3924}\\
Traffic&24&0.8557&0.8277&1.1204&\textbf{0.8163}\\
&48&0.8691&\textbf{0.8537}&1.0889&0.8609\\
&96&0.8751&0.8680&1.0919&\textbf{0.8634}\\
Exchange&24&0.0587&0.0568&\textbf{0.0443}&0.0466\\
&48&0.1110&0.1078&\textbf{0.0894}&0.0930\\
&96&0.2120&0.2086&\textbf{0.1999}&0.2012\\
Solar&24&0.3050&0.2921&0.8833&\textbf{0.2034}\\
&48&0.5253&0.4803&1.1849&\textbf{0.3529}\\
&96&0.6662&0.6102&1.3160&\textbf{0.4713}\\
\bottomrule
\end{tabular}
\end{adjustbox}
\end{table}

\subsection{Conventional-Similarity Diagnostic for the Fixed-Support Ablation}
Pearson is centered cosine and therefore induces the Pearson ranking for these univariate windows. Raw cosine retains level information. Last-value-anchored L2 subtracts the final observed value from each past window before measuring Euclidean distance, so it compares the shape of the recent trajectory relative to its endpoint. Spectral cosine compares log-magnitude real fast Fourier transform (rFFT) features after mean removal. Unlike the restricted \prstat ablation used in this diagnostic, these conventional rules search the full temporally admissible same-channel memory, giving them favorable access to the candidate set. This experiment tests whether the fixed-support \prstat gains can be explained by a weak Pearson baseline; it is not a head-to-head comparison between these rules and full \prr.

\begin{table}[h]
\caption{Conventional-similarity diagnostic for the fixed-support ablation (AnalogFutureMSE $\downarrow$). Bold is the best entry per row among the displayed retrieval rules.}
\label{tab:app_similarity}
\centering
\scriptsize
\begin{adjustbox}{max width=\linewidth}
\begin{tabular}{llrrrrr}
\toprule
Dataset & $H$ & Raw cos. & Pearson & Anchored L2 & Spectral & \prstat\\
\midrule
Electricity&24&0.3909&0.4138&\textbf{0.2918}&0.5877&0.3360\\
&48&0.4252&0.4557&\textbf{0.3214}&0.6325&0.3753\\
&96&0.4495&0.4771&\textbf{0.3358}&0.7060&0.3924\\
Traffic&24&0.8585&0.8557&0.8208&1.1801&\textbf{0.8163}\\
&48&0.8726&0.8691&\textbf{0.8449}&1.1234&0.8609\\
&96&0.8736&0.8751&\textbf{0.8486}&1.1489&0.8634\\
Exchange&24&0.0561&0.0587&\textbf{0.0433}&0.0550&0.0466\\
&48&0.1048&0.1110&\textbf{0.0855}&0.1041&0.0930\\
&96&0.2165&0.2120&\textbf{0.1817}&0.2027&0.2012\\
Solar&24&0.3680&0.3050&0.3114&0.6766&\textbf{0.2034}\\
&48&0.5418&0.5253&0.5004&2.1285&\textbf{0.3529}\\
&96&0.5915&0.6662&0.5334&2.9222&\textbf{0.4713}\\
\bottomrule
\end{tabular}
\end{adjustbox}
\end{table}

\prstat is lower than raw cosine, Pearson, and spectral cosine in all 12 tasks. Against anchored L2, it is lower in 4/12 tasks; the only condition-wise significant \prstat-over-L2 gains are the three Solar horizons, while the condition-wise tests favor L2 significantly on Electricity and Exchange. Thus the fixed-support ablation does not universally dominate strong handcrafted similarities, particularly anchored L2. This does not alter the main full-\prr claim, which is framed as learning predictive relevance rather than universal dominance over every retrieval metric.

\subsection{Long-Horizon Anchored-L2 Stress Test}
The short-horizon result above motivates a stronger deterministic stress test on the exact 24 frozen long-horizon manifests used for Table~\ref{tab:main}. Anchored L2 uses the same-channel admissible memory, input lookback $L=96$, memory stride 8, and Top-10 retrieval, but searches the full admissible memory and requires no training or future supervision. Table~\ref{tab:l2_long_retrieval} summarizes the matched comparison.

\begin{table}[h]
\caption{Long-horizon anchored-L2 stress test on the 24 frozen retrieval conditions. L2 gain vs. Pearson is the mean AnalogFutureMSE reduction across four horizons. PRR wins counts horizons where full \prr has lower AnalogFutureMSE than anchored L2; mean PRR gain vs. L2 is positive when \prr is better.}
\label{tab:l2_long_retrieval}
\centering
\scriptsize
\setlength{\tabcolsep}{5.0pt}
\begin{tabular}{lrrr}
\toprule
Dataset & L2 gain vs. Pearson & PRR wins vs. L2 & Mean PRR gain vs. L2\\
\midrule
ETTh1 & +28.20\% & 2/4 & +2.70\%\\
Weather & +37.10\% & 2/4 & +15.96\%\\
Electricity & +38.55\% & 1/4 & -2.01\%\\
Traffic & +29.43\% & 0/4 & -6.39\%\\
Exchange & +8.67\% & 2/4 & -2.69\%\\
Solar & +15.55\% & 0/4 & -4.93\%\\
\midrule
Overall & +26.25\% & 7/24 & +0.44\%\\
\bottomrule
\end{tabular}
\end{table}

Anchored L2 improves over Pearson in all 24 conditions, with a 26.25\% mean and 27.56\% median reduction in AnalogFutureMSE. Full \prr is better than anchored L2 in 7/24 conditions; its mean relative gain is +0.44\%, while the median is -4.28\%. The comparison is strongly domain- and horizon-dependent: \prr has large advantages on Weather at $H=96$ and $192$, whereas anchored L2 is consistently stronger on Traffic and Solar. The Pearson-to-\prr result therefore establishes improvement over the Pearson baseline, not universal superiority to every deterministic similarity rule.

\subsection{Matched Recent Retrieval Baseline: CRAFT}
\label{app:craft_matched}
To address whether the Pearson comparison is overly weak, we implement the retrieval component of CRAFT \citep{kang2026craft} under the same frozen long-horizon protocol. The matched setting keeps lookback $L=96$, memory stride 8, the exact query--channel manifests, and Top-10 AnalogFutureMSE evaluation fixed. Following CRAFT's frequency-ratio design, we retain the lowest $F=5$ complex Fourier coefficients (5\% of $L$). We report two variants: \emph{CRAFT-SC@10}, which applies the frequency score within the same-channel admissible memory for an apples-to-apples scoring-rule comparison, and \emph{CRAFT-Graph@10}, which first builds CRAFT's train-only top-$M=3$ channel-similarity graph and retrieves from those graph channels. The latter is closer to CRAFT's channel-wise retrieval mechanism; only 42.1\% of its final Top-10 candidates are same-channel on average. This experiment isolates retrieval selection only: it does not reproduce CRAFT's direct forecaster, reference projection, additive integration coefficient, or end-to-end forecasting benchmark.

\begin{table}[h]
\caption{Matched CRAFT retrieval-rule comparison (AnalogFutureMSE $\downarrow$). The primary scope is the 16 conditions from the four datasets shared with CRAFT's reported benchmark; the full 24-condition rows are secondary because Exchange and Solar are outside that overlap. W/T/L and mean/median gain are for the left-named method relative to the right-named comparator.}
\label{tab:craft_matched}
\centering
\scriptsize
\setlength{\tabcolsep}{3.4pt}
\begin{adjustbox}{max width=\linewidth}
\begin{tabular}{llrrr}
\toprule
Scope & Comparison & W/T/L & Mean gain & Median gain\\
\midrule
CRAFT-overlap 16 & CRAFT-SC@10 vs. Pearson & 4/0/12 & -4.27\% & -5.40\%\\
 & CRAFT-Graph@10 vs. Pearson & 11/0/5 & +1.84\% & +1.33\%\\
 & \prr vs. CRAFT-SC@10 & 16/0/0 & +37.80\% & +39.80\%\\
 & \prr vs. CRAFT-Graph@10 & 16/0/0 & +33.68\% & +36.34\%\\
\midrule
All 24 & CRAFT-Graph@10 vs. Pearson & 16/0/8 & +2.51\% & +1.88\%\\
 & \prr vs. CRAFT-Graph@10 & 18/0/6 & +23.84\% & +29.41\%\\
\bottomrule
\end{tabular}
\end{adjustbox}
\end{table}

On the CRAFT-overlap scope, CRAFT-Graph is a stronger baseline than Pearson in 11/16 conditions, yet \prr has lower AnalogFutureMSE in all 16. The same-channel frequency-only variant is weaker than Pearson on average, confirming that CRAFT's cross-channel graph contributes materially under this matched protocol. Over all six datasets, CRAFT-Graph beats Pearson in 16/24 conditions and beats \prr in 6/24; those additional wins arise outside the four-dataset overlap and reinforce the broader finding that retrieval inductive biases remain domain dependent. The result therefore strengthens the baseline comparison without implying an end-to-end ranking of the original forecasting systems.

\subsection{Matched Canonical Retrieval Baseline: RAFT}
\label{app:raft_matched}
We additionally adapt the public RAFT retrieval rule \citep{han2025raft} to the same frozen long-horizon benchmark. RAFT decomposes each multivariate past window at temporal granularities $g\in\{4,2,1\}$, removes the endpoint offset, and retrieves by normalized correlation of the flattened multivariate representation. Following the public implementation, each scale first keeps Top-20 candidates and uses temperature 0.1. Because the original forecasting architecture consumes separate scale-specific retrievals rather than defining one unique Top-10 history list, we report two transparent matched projections. \emph{RAFT-MS@10} averages the official Top-20 softmax probability mass across scales and selects the resulting Top-10 histories. \emph{RAFT-ValScale@10} is a stronger control that selects $g\in\{1,2,4\}$ using validation AnalogFutureMSE only and freezes that scale on test. Both variants use the same $L=96$, stride-8 temporally admissible memory, frozen query--channel manifests, and Top-10 AnalogFutureMSE metric as \prr. This is a retrieval-rule comparison, not a reproduction of RAFT's end-to-end forecasting benchmark.

\begin{table}[h]
\caption{Matched RAFT retrieval-rule comparison on all 24 frozen conditions (AnalogFutureMSE $\downarrow$). W/T/L and mean/median gain are for the left-named method relative to the right-named comparator. RAFT-ValScale chooses the temporal granularity using validation only and freezes it on test.}
\label{tab:raft_matched}
\centering
\scriptsize
\setlength{\tabcolsep}{4.0pt}
\begin{adjustbox}{max width=\linewidth}
\begin{tabular}{lrrr}
\toprule
Comparison & W/T/L & Mean gain & Median gain\\
\midrule
RAFT-MS@10 vs. Pearson & 13/0/11 & +7.57\% & +1.12\%\\
RAFT-ValScale@10 vs. Pearson & 15/0/9 & +7.90\% & +1.61\%\\
\prr\ vs. RAFT-MS@10 & 22/0/2 & +20.58\% & +21.01\%\\
\prr\ vs. RAFT-ValScale@10 & 22/0/2 & +20.34\% & +20.68\%\\
\bottomrule
\end{tabular}
\end{adjustbox}
\end{table}

The validation-selected RAFT control improves Pearson in 15/24 conditions, confirming that it is a materially stronger similarity-based baseline under this protocol. Nevertheless, \prr has lower AnalogFutureMSE in 22/24 conditions against both RAFT projections; the two RAFT wins occur on Exchange at $H=192$ and $H=336$. Validation selects $g=1$ in 15 conditions, $g=2$ in five, and $g=4$ in four, so the result is not tied to a single temporal granularity. RAFT-MS also outperforms CRAFT-Graph in 14/24 conditions under the matched protocol, providing an additional check that the external baseline set is not uniformly weak. As with CRAFT, these results rank retrieval rules under our frozen relevance metric rather than the complete original forecasting systems.

\subsection{Downstream Anchored-L2 Integration Stress Test}
We next attach anchored-L2 retrieval to the same five frozen forecasters using the identical seven-point validation-only trust grid as full \prr. No forecaster is retrained and no test outcome is used to select trust. Anchored L2 with condition-specific calibration improves 82/120 conditions, ties 14, and degrades 24, with a +1.855\% mean MSE gain over Direct. Full \prr gives 79/15/26 and +1.407\% under the same protocol. Table~\ref{tab:l2_downstream_pair} reports the exact paired comparison.

\begin{table}[h]
\caption{Exact paired downstream comparison between validation-calibrated anchored L2 and full \prr on the same 120 frozen conditions. L2-vs.-PRR W/T/L counts conditions where calibrated anchored L2 has lower/equal/higher MSE. Mean gains vs. Direct are averaged over the same 20 conditions per dataset; the final column is positive when anchored L2 is better than \prr.}
\label{tab:l2_downstream_pair}
\centering
\scriptsize
\setlength{\tabcolsep}{4.0pt}
\begin{adjustbox}{max width=\linewidth}
\begin{tabular}{lrrrr}
\toprule
Dataset & L2 vs. PRR W/T/L & L2 gain vs. Direct & PRR gain vs. Direct & Mean L2 gain vs. PRR\\
\midrule
ETTh1 & 16/1/3 & -0.12\% & -0.75\% & +0.62\%\\
Weather & 3/0/17 & +1.35\% & +2.39\% & -1.06\%\\
Electricity & 7/2/11 & +0.37\% & +0.40\% & -0.02\%\\
Traffic & 5/6/9 & +0.48\% & +0.51\% & -0.03\%\\
Exchange & 16/1/3 & +2.12\% & +0.35\% & +1.85\%\\
Solar & 17/0/3 & +6.92\% & +5.55\% & +1.42\%\\
\midrule
Overall & 64/10/46 & +1.85\% & +1.41\% & +0.46\%\\
\bottomrule
\end{tabular}
\end{adjustbox}
\end{table}

The paired aggregate difference is small: anchored L2 has a +0.463\% mean and +0.070\% median MSE advantage over full \prr. The dataset structure is more informative than the global average. Anchored L2 is stronger on Exchange and Solar and reduces the ETTh1 failure, whereas full \prr is clearly stronger on Weather; Electricity and Traffic are nearly tied in mean downstream gain. This reinforces the relevance--utility interpretation: a learned relevance model and a strong task-aligned handcrafted similarity can induce different but comparably useful retrieval signals, and neither rule dominates across domains.

\subsection{Strong-Base Retriever Substitution}
The anchored-L2 stress test raises a natural question: is full \prr tied to Pearson as its deterministic candidate branch, or can the same learned machinery benefit from a stronger handcrafted base? We therefore form \emph{Anchored-PRR} (A-PRR) by replacing Pearson Top-100 with anchored-L2 Top-100 while retaining the frozen future-supervised learned-representation Top-100 branch from the original full-\prr stack. The two Top-100 sets are unioned exactly as in full \prr. Only the fusion reranker is retrained with the same future-supervised listwise-plus-pairwise objective; the learned encoders are not retrained. Anchored-L2 ranking is represented to the residual reranker by a query-wise standardized negative anchored MSE, which preserves the exact anchored-L2 candidate ordering.

\begin{table}[h]
\caption{Strong-base substitution ablation on the 24 frozen long-horizon retrieval conditions. A-PRR replaces the Pearson candidate branch with anchored L2 while retaining the frozen learned branch. W/L counts are over four horizons per dataset; positive mean gain means A-PRR has lower AnalogFutureMSE. The final column reports the fraction of A-PRR's final Top-10 selections that come from learned-only candidates outside anchored-L2 Top-100.}
\label{tab:aprr_strong_base}
\centering
\scriptsize
\setlength{\tabcolsep}{3.8pt}
\begin{adjustbox}{max width=\linewidth}
\begin{tabular}{lrrrrr}
\toprule
Dataset & A-PRR vs. L2 W/L & Mean gain vs. L2 & A-PRR vs. PRR W/L & Mean gain vs. PRR & Learned-only Top-10\\
\midrule
ETTh1 & 3/1 & +4.12\% & 3/1 & +1.47\% & 66.4\%\\
Weather & 2/2 & +18.20\% & 3/1 & +1.38\% & 48.9\%\\
Electricity & 0/4 & -2.18\% & 3/1 & -0.18\% & 19.1\%\\
Traffic & 0/4 & -2.90\% & 4/0 & +3.26\% & 21.0\%\\
Exchange & 1/3 & -2.57\% & 3/1 & -0.96\% & 78.9\%\\
Solar & 0/4 & -4.86\% & 3/1 & +0.07\% & 36.9\%\\
\midrule
Overall & 6/18 & +1.63\% & 19/5 & +0.84\% & 45.2\%\\
\bottomrule
\end{tabular}
\end{adjustbox}
\end{table}

A-PRR improves the original Pearson-based \prr in 19/24 conditions, with a +0.84\% mean and +1.37\% median relative gain, and it improves Pearson itself in 24/24 conditions with a +27.51\% mean reduction. The stronger deterministic branch therefore improves the learned framework in most conditions, showing that \prr is not intrinsically tied to Pearson. However, A-PRR surpasses standalone anchored L2 in only 6/24 conditions: the mean gain is +1.63\% but the median is -2.39\%, indicating that a few large gains---especially on Weather---raise the aggregate mean while anchored L2 remains better in most individual conditions. Support expansion is nevertheless active rather than ignored: the mean A-PRR union contains 173.8 candidates, including 73.8 learned-only candidates outside anchored-L2 Top-100, and 45.2\% of the final Top-10 selections come from those learned-only candidates. Thus stronger base retrieval can improve \prr, but learned support expansion and reranking do not uniformly improve an already strong task-aligned deterministic rule.

\subsection{Candidate-Pool Coverage Diagnostic}
To test whether the full-memory L2 advantage is caused by access to candidates absent from Pearson Top-100, we compute
\begin{equation}
\mathrm{Coverage@10}(q)=\frac{|\mathrm{Top10}_{\mathrm{L2,full}}(q)\cap\mathrm{Top100}_{\mathrm{Pearson}}(q)|}{10}.
\end{equation}
We also rerank the \emph{same} Pearson Top-100 (P100) using anchored L2 (L2-within-P100). Table~\ref{tab:app_coverage} shows that candidate-pool access alone does not explain the domain-dependent L2 advantage.

\begin{table}[h]
\caption{Candidate-pool diagnostic. Pool restriction $\Delta$ is the relative change in L2 AnalogFutureMSE when restricted to Pearson Top-100; positive values mean the restriction worsens L2 and negative values mean it improves L2. PRR-Stat gain vs. P100-L2 is positive when \prstat is better.}
\label{tab:app_coverage}
\centering
\scriptsize
\begin{adjustbox}{max width=\linewidth}
\begin{tabular}{llrrrrr}
\toprule
Dataset & $H$ & Coverage@10 & Full L2 & L2-within-P100 & Pool restriction $\Delta$ & \prstat gain vs. P100-L2\\
\midrule
Electricity&24&91.1\%&0.2918&0.2964&+1.6\%&-13.4\%\\
&48&91.2\%&0.3214&0.3268&+1.7\%&-14.8\%\\
&96&91.2\%&0.3358&0.3429&+2.1\%&-14.4\%\\
Traffic&24&97.7\%&0.8208&0.8189&-0.2\%&+0.3\%\\
&48&97.8\%&0.8449&0.8417&-0.4\%&-2.3\%\\
&96&97.9\%&0.8486&0.8453&-0.4\%&-2.1\%\\
Exchange&24&84.6\%&0.0433&0.0427&-1.4\%&-9.2\%\\
&48&84.6\%&0.0855&0.0836&-2.2\%&-11.3\%\\
&96&85.0\%&0.1817&0.1772&-2.5\%&-13.6\%\\
Solar&24&89.0\%&0.3114&0.2795&-10.2\%&+27.2\%\\
&48&89.3\%&0.5004&0.4415&-11.8\%&+20.1\%\\
&96&89.4\%&0.5334&0.4848&-9.1\%&+2.8\%\\
\bottomrule
\end{tabular}
\end{adjustbox}
\end{table}

On Electricity, restricting L2 to Pearson Top-100 causes only a 1.6--2.1\% penalty, and L2 remains significantly better than \prstat under the condition-wise test at every horizon. On Exchange, restriction slightly \emph{improves} L2, which also remains significantly better under the condition-wise test. On Solar, restriction improves L2 as well, but \prstat remains better, with condition-wise significant gains at $H=24$ and $H=48$. Thus the observed domain contrast persists even when the candidate pool is held fixed.

\section{Additional Robustness and Scope Notes}
\subsection{Scope of the Full-PRR Claim}
\label{app:scope}
The central claim is not that \prr universally dominates every past-only similarity rule, but that future supervision can learn a deployable past-only relevance function that substantially improves a conventional similarity baseline. The \prbudget control shows that this conclusion does not depend on a larger test-time candidate budget, while the matched RAFT, CRAFT, anchored-L2, and strong-base substitution experiments characterize settings in which alternative retrieval inductive biases remain competitive. The broader design space remains open: alternative relevance utilities may induce different rankings, full-memory search may differ from the union of two Top-100 supports, the learned encoder and reranker are dataset--horizon specific rather than zero-shot universal, and downstream trust is condition-level rather than query-adaptive. These limitations motivate stronger retrieval inductive biases and forecast-utility-aware integration without changing the retrieval-learning contribution studied here.

\subsection{Cross-Channel vs. Same-Channel Interpretation}
The controlled six-benchmark mechanism study uses same-channel candidates to ensure semantic comparability of future trajectories. This should not be interpreted as a claim that cross-channel retrieval is undesirable. Complementary cross-channel experiments on Electricity and Traffic use the restricted \prstat ablation and indicate that cross-channel candidate pools can also be informative, but they are diagnostics rather than evidence for the full \prr method. Candidate admissibility is therefore part of the retrieval problem specification rather than a universal modeling choice.

\subsection{Scope of the Relevance Regime Map}
The coordinates in the \prstat relevance-regime analysis are empirical diagnostics under its fixed-pool representation, candidate generator, and control construction. They are not intrinsic dataset constants and may move with richer observable features or different utilities. What is robust in our experiments is the qualitative existence of sharply contrasting cases: candidate priors dominate Weather and Exchange, whereas preserving correct query--future correspondence is decisive for ETTh1, Traffic, and Solar. Electricity demonstrates that both effects can coexist.

\section{Seed-Level and Frozen-Evaluation Diagnostics}
\label{app:seed_diagnostics}
The tables in this section use $H\in\{24,48,96\}$ for seed-level and mechanism diagnostics of \prstat and its controls. The five-seed stability suite is restricted to Electricity, Traffic, Exchange, and Solar; ETTh1 and Weather use the separate three-seed mechanism protocol in Appendix D.5. The aligned long-horizon Pearson-vs.-\prr results are reported in Table~\ref{tab:main}; \prstat is used here only for fixed-support diagnostics.
\subsection{Five-Seed Stability of PRR-Stat and Shuffled Models}
\label{app:seed_stability}
Table~\ref{tab:seed_full} reports the complete five-seed mean and coefficient of variation for the four-dataset short-horizon fixed-support suite. \prstat remains stable across all four datasets. The Shuffled model is more variable on Electricity, but this variability does not explain the \prstat-vs.-Shuffled ordering because \prstat is better at every seed and horizon there. Exchange provides the opposite case: \prstat and Shuffled have comparable means and low variance, consistent with no detectable \prstat-over-Shuffled advantage under this diagnostic.

\begin{table}[h]
\caption{Five-seed AnalogFutureMSE stability. The coefficient of variation (CV) is standard deviation divided by mean.}
\label{tab:seed_full}
\centering
\scriptsize
\begin{tabular}{llrrrrrr}
\toprule
Dataset & Method & \multicolumn{2}{c}{$H=24$} & \multicolumn{2}{c}{$H=48$} & \multicolumn{2}{c}{$H=96$}\\
 & & Mean & CV\% & Mean & CV\% & Mean & CV\%\\
\midrule
Electricity & \prstat & 0.3360 & 0.66 & 0.3753 & 0.83 & 0.3924 & 0.60\\
 & Shuffled & 0.4381 & 2.78 & 0.5025 & 6.55 & 0.5545 & 7.67\\
Traffic & \prstat & 0.8163 & 1.13 & 0.8609 & 1.24 & 0.8634 & 1.19\\
 & Shuffled & 1.1204 & 1.18 & 1.0889 & 1.98 & 1.0919 & 1.53\\
Exchange & \prstat & 0.0466 & 2.72 & 0.0930 & 1.43 & 0.2012 & 1.04\\
 & Shuffled & 0.0443 & 2.50 & 0.0894 & 5.66 & 0.1999 & 1.84\\
Solar & \prstat & 0.2034 & 2.19 & 0.3529 & 2.41 & 0.4713 & 1.43\\
 & Shuffled & 0.8833 & 4.97 & 1.1849 & 4.83 & 1.3160 & 1.09\\
\bottomrule
\end{tabular}
\end{table}

\subsection{Task-Level Significance Results}
All significance statements in this section are condition-wise moving-block bootstrap diagnostics and should not be interpreted as simultaneous family-wise claims. Across the 12 short-horizon fixed-support tasks, \prstat outperforms Pearson on 12/12 and the improvement is condition-wise significant on 9/12. \prstat outperforms Shuffled with condition-wise significance at all Electricity, Traffic, and Solar horizons (9/9), while it does not outperform Shuffled at any Exchange horizon. Candidate Prior yields condition-wise significant improvements over Pearson at all Electricity and Exchange horizons (6/6) and at none of the Traffic or Solar horizons. Table~\ref{tab:sigpattern} records this pattern explicitly.

\begin{table}[h]
\caption{Task-level mechanism evidence in the short-horizon frozen evaluation suite. Under the condition-wise moving-block bootstrap, ``sig.'' denotes statistically significant and ``n.s.'' denotes not significant.}
\label{tab:sigpattern}
\centering
\scriptsize
\begin{tabular}{lccccl}
\toprule
Dataset & $H$ & Stat$<$Pearson & Stat$<$Shuffled & Prior$<$Pearson & Regime\\
\midrule
Electricity&24&sig.&sig.&sig.&query-specific + global\\
&48&sig.&sig.&sig.&query-specific + global\\
&96&sig.&sig.&sig.&query-specific + global\\
Traffic&24&sig.&sig.&no&query-specific\\
&48&yes, n.s.&sig.&no&query-specific\\
&96&yes, n.s.&sig.&no&query-specific\\
Exchange&24&sig.&no&sig.&candidate-global\\
&48&sig.&no&sig.&candidate-global\\
&96&yes, n.s.&no&sig.&candidate-global\\
Solar&24&sig.&sig.&no&query-specific\\
&48&sig.&sig.&no&query-specific\\
&96&sig.&sig.&no&query-specific\\
\bottomrule
\end{tabular}
\end{table}

\section{Additional Mechanism Diagnostics}
\subsection{Target-Scaling Robustness on Weather}
Table~\ref{tab:weather_scale} records the three stages of the retrospective Weather target-scaling diagnosis. Under the local definition, the first quartile of the test-window past standard deviation was approximately 0.0066, compared with a train-memory 10th-percentile reference of approximately 0.0345, and the maximum absolute local target magnitude reached approximately 130--138. Dividing by such small query-window scales can therefore produce extreme target magnitudes and AnalogFutureMSE values. A train-memory quantile floor reduces but does not eliminate this instability. The final fixed train-channel scale removes the query-window denominator and yields numerically comparable targets; under this frozen definition, the qualitative candidate-global pattern in the Weather \prstat mechanism diagnostic remains. This subsection is a retrospective scaling diagnostic, not an additional full-\prr performance claim; all aligned long-horizon comparisons elsewhere use only the final fixed train-channel target definition.
\begin{table}[h]
\caption{Retrospective Weather target-scaling diagnostic using Pearson retrieval (AnalogFutureMSE $\downarrow$). The aligned benchmark uses only the final fixed train-channel definition.}
\label{tab:weather_scale}
\centering
\scriptsize
\begin{tabular}{lrrr}
\toprule
Target & $H=24$ & $H=48$ & $H=96$\\
\midrule
Local past std & 140.7876 & 199.0089 & 1460.3328\\
Local std + training 10th-percentile floor & 24.3401 & 22.7755 & 30.4483\\
Fixed train-channel scale (Same) & 0.3963 & 0.7132 & 0.9131\\
\bottomrule
\end{tabular}
\end{table}

\subsection{Deconfounding as an Identification Experiment}
Table~\ref{tab:deconf} shows that subtracting an empirical candidate prior from the future-distance target does not improve the fixed-support \prstat retriever. On ETTh1, a clear \prstat-vs.-Shuffled gap remains after deconfounding, which is consistent with residual query-specific signal under this diagnostic; however, absolute AnalogFutureMSE becomes substantially worse than for raw \prstat. On Weather, deconfounding does not restore a \prstat advantage over the corresponding Shuffled control and also worsens absolute error. We therefore use prior subtraction only as an identification/mechanism diagnostic. It is neither the retrieval objective of full \prr nor evidence that the population decomposition in Eq.~(14) prescribes an explicit $G+H$ architecture.
\begin{table}[h]
\caption{Raw and candidate-prior-deconfounded fixed-support mechanism diagnostic (AnalogFutureMSE $\downarrow$). ``Stat'' denotes \prstat and ``Shuf'' denotes the matched Shuffled-Future control.}
\label{tab:deconf}
\centering
\scriptsize
\begin{tabular}{llrrrrrr}
\toprule
Data & $H$ & Pearson & Raw Stat & Raw Shuf & Prior & Deconf Stat & Deconf Shuf\\
\midrule
ETTh1&24&0.9308&0.7820&1.0593&1.1888&1.0464&1.3239\\
&48&1.0527&0.8679&1.2434&1.1882&1.1733&1.4947\\
&96&1.1959&0.9954&1.3874&1.2183&1.3540&1.7204\\
Weather&24&0.3963&0.3565&0.3150&0.1375&0.7663&0.7124\\
&48&0.7132&0.5772&0.5607&0.1988&1.1009&1.0569\\
&96&0.9131&0.8063&0.6277&0.2579&1.4818&1.2679\\
\bottomrule
\end{tabular}
\end{table}

\end{document}